\documentclass[aip, amsmath, amssymb, reprint,]{revtex4-1}

\usepackage{graphicx}
\usepackage{dcolumn}
\usepackage{bm}

\usepackage{xcolor}
\usepackage{units}
\usepackage[utf8]{inputenc}
\usepackage[T1]{fontenc}
\usepackage{mathptmx}
\usepackage{etoolbox}
\usepackage{subcaption}
\usepackage{stackengine}
\usepackage{caption}
\makeatletter
\def\@email#1#2{
 \endgroup
 \patchcmd{\titleblock@produce}
  {\frontmatter@RRAPformat}
  {\frontmatter@RRAPformat{\produce@RRAP{*#1\href{mailto:#2}{#2}}}\frontmatter@RRAPformat}
  {}{}
}
\makeatother
\begin{document}

\preprint{}

\title{A New Optimization Methodology for Polar Direct Drive Illuminations at the National Ignition Facility}
\author{D.E.M. Barlow}
\affiliation{CELIA, Universit\'{e} de Bordeaux, Talence 33400, France}

\author{A. Cola\"{i}tis}
\affiliation{CELIA, Universit\'{e} de Bordeaux, Talence 33400, France}
\affiliation{Laboratory for Laser Energetics, University of Rochester, Rochester, New York 14623-1299, USA}

\author{D. Viala}
\affiliation{CELIA, Universit\'{e} de Bordeaux, Talence 33400, France}

\author{M. J. Rosenberg}
\author{I. Igumenshchev}
\author{V. Goncharov}
\author{L. Ceurvorst}
\author{P. B. Radha}
\author{W. Theobald}
\author{R. S. Craxton}
\affiliation{Laboratory for Laser Energetics, University of Rochester, Rochester, New York 14623-1299, USA}

\author{M.J.V. Streeter}
\affiliation{School of Mathematics and Physics, Queen’s University Belfast, Belfast BT7 1NN, United Kingdom}

\author{R. H. H. Scott}
\author{K. Glize}
\affiliation{Central Laser Facility, STFC Rutherford Appleton Laboratory, Harwell Oxford, OX11 0QX, United Kingdom}

\author{T. Chapman}
\affiliation{Lawrence Livermore National Laboratory, Livermore, California 94550, USA}

\author{J. Mathiaud}
\affiliation{CELIA, Universit\'{e} de Bordeaux, Talence 33400, France}

\email{Duncan.Barlow@u-bordeaux.fr}
\thanks{}

\date{\today}

\begin{abstract}
A new, efficient, algorithmic approach to create illumination configurations for laser driven high energy density physics experiments is proposed. The method is applied to a polar direct drive solid target experiment at the National Ignition Facility (NIF), where it is simulated to create more than $\times 2$ higher peak pressure and $\times 1.4$ higher density by maintaining better shock uniformity. The analysis is focused on projecting shocks into solid targets at the NIF, but with minor adaptations the method could be applied to implosions, other target geometries and other facilities.
\end{abstract}

\maketitle
\def\stackalignment{l}

\begin{figure*}
\centering
\begin{subfigure}[t]{0.33\textwidth}
  \centering
  \topinset{(a)}{\includegraphics[width=0.9\linewidth,trim=0mm 0mm 0mm 0mm,clip]{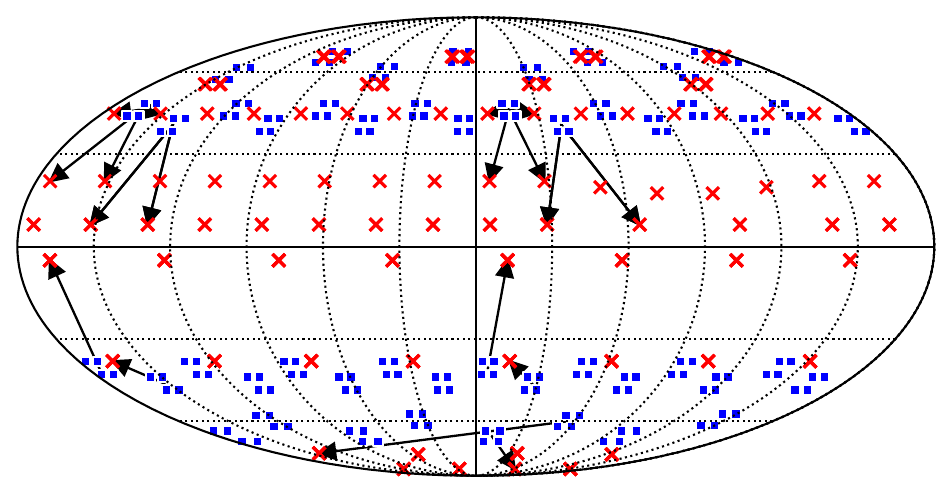}}{0pt}{0pt}
\end{subfigure}
\centering
\begin{subfigure}[t]{0.33\textwidth}
  \centering
  \topinset{(b)}{\includegraphics[width=1.0\linewidth,trim=0mm 0mm 0mm 0mm,clip]{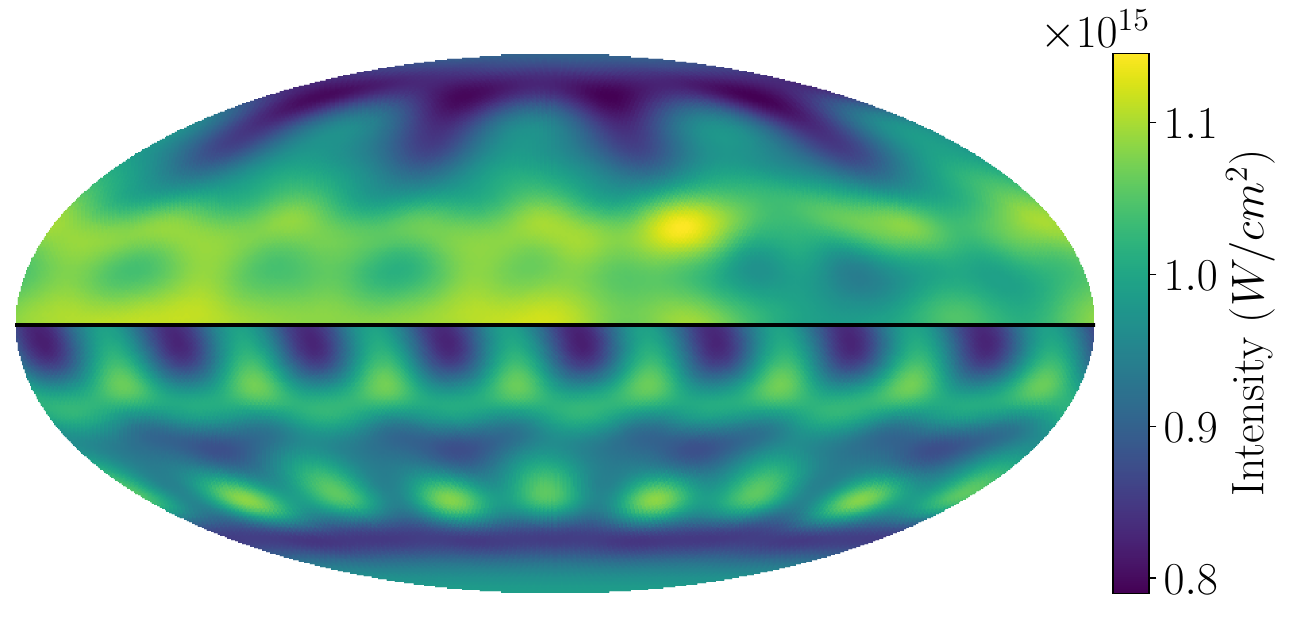}}{0pt}{0pt}
\end{subfigure}
\centering
\begin{subfigure}[t]{0.33\textwidth}
  \centering
  \topinset{(c)}{\includegraphics[width=1.0\linewidth,trim=0mm 0mm 0mm 0mm,clip]{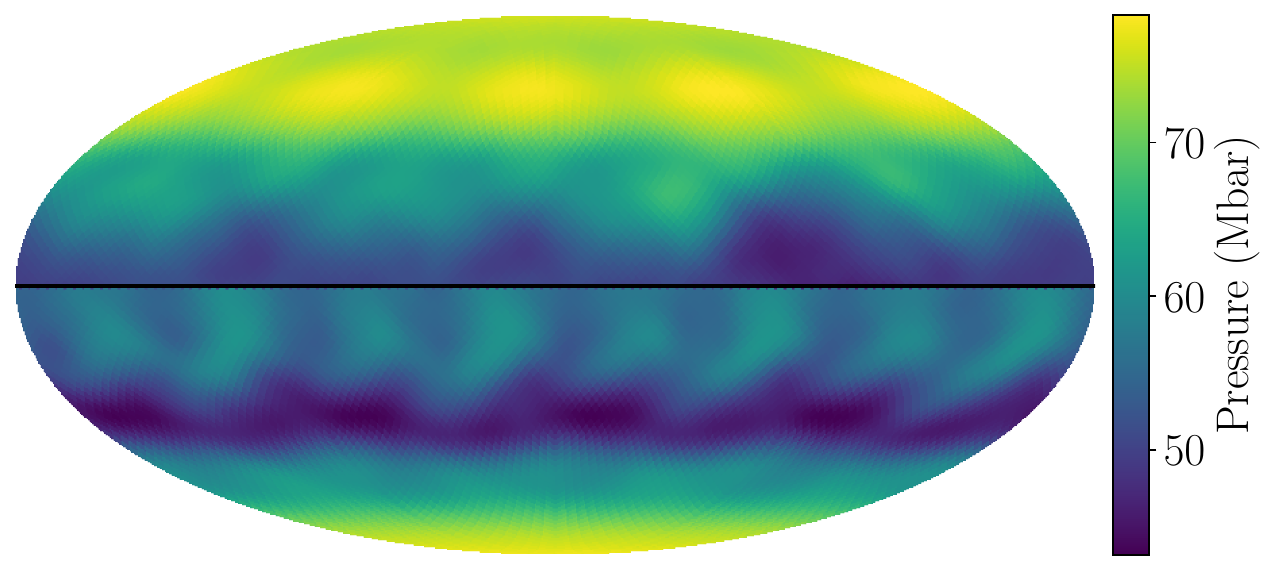}}{0pt}{0pt}
\end{subfigure}
\caption{ (a) Mollweide projections showing the NIF target chamber beam locations (blue squares) linked with black arrows to the beam pointing intersections with a 1100 $\mathrm{\mu m}$ radius target (red crosses). The layouts are symmetric about the equator and so half is shown of: (top) A common PDD approach, used in N190204-003 and (bottom) the ``Optimized Config" (OC) generated with the method presented in this Letter. In (b) and (c) Mollweide projections each show the target illuminated with N190204-003 (top) and OC (bottom). (b) An illumination without an ablation plasma, and (c) is simulated with a plasma at $4 \mathrm{ns}$ through the laser pulse, with the impacts of CBET and converted into an approximation of ablation pressure using Equation \ref{eqn:densityofabsorption}.}
\label{fig:mollweide_pointings}
\end{figure*}

The highest energy laser facilities in the world, the National Ignition Facility (NIF) \cite{miller2004national} and Laser Mega-joule (LMJ) \cite{miquel2016laser}, are configured with beam ports in the polar regions of the target chamber, for indirect drive inertial confinement fusion (ICF) \cite{nuckolls1972laser,atzeni2004physics}. The laser energy is incident on the inside of a cylindrical target, where it is converted to x-rays. The x-rays drive an implosion capsule to achieve ignition conditions \cite{lindl2004physics}. The thermal x-ray bath helps to maintain uniform drive throughout the implosion, which is one of the most significant challenges in ICF \cite{li2004effects}. Using indirect drive, the NIF recently achieved thermonuclear ignition \cite{abu2022lawson}. To move into the regime of energy production, fusion energy output must be increased, which requires coupling more laser energy to the target. Direct drive (DD) \cite{craxton2015direct} achieves higher laser-to-target coupling, but with stringent conditions on laser driver uniformity. DD facilities are now capable of attaining $\sigma < 2\%$ deviations in drive symmetry required for high performance implosions \cite{colaitis2022prl} however their laser energy is too small to probe ignition conditions \cite{boehly1997omega,nora2014theory}.

Polar direct drive (PDD) is used to carry out DD experiments at the mega-joule laser facilities \cite{skupsky2004polar,marozas2006polar,canaud2007high,yeamans2021high}. The configurations are numerous and varied, but repointing of the laser beams toward the target equator is often used to distribute energy more uniformly. Due to the large laser energies available, PDD has proven to be a useful technique for exploring high energy density physics \cite{regan2018national,ceurvorst2022development,viala2024cbetsolid}, laser-plasma instabilities (LPI) \cite{marozas2018first,solodov2022hot,barlow2022role}, hydrodynamic scaling \cite{nora2014theory,murphy2015laser,regan2018national,campbell2021direct,rosenberg2023hot} and reliable neutron production \cite{zylstra2020enhanced,yeamans2021high}. However, high performance ICF implosions are currently beyond the reach of PDD at the NIF due to several technical challenges including, laser imprint, DD cryogenic target positioner, and issues maintaining uniform drive. Instead of gas filled implosions capsules, solid plastic targets (often doped or deuterated) can provide an easy-to-diagnose platform for laser-target coupling experiments in PDD \cite{ceurvorst2022development,viala2024cbetsolid,barlow2022role} and the illuminations are still relevant for implosion targets.

Cross-beam energy transfer (CBET) is an LPI that significantly modifies laser coupling and drive distribution for indirect drive \cite{michel2009tuning}, and DD \cite{igumenshchev2010crossed}, including PDD \cite{murphy2015laser,marozas2018first}. It is a type of stimulated Brillouin scattering due to resonance between two laser waves and an ion-acoustic wave in the plasma. It is also the most important LPI for drive uniformity at the intensities  $ 10^{14}<I<10^{15} \mathrm{W/cm^2}$ and laser wavelength $\lambda = 351 \mathrm{nm}$ which are currently the focus of DD. CBET is one of the few LPI that can be effectively modelled with ray-tracing while coupled to a radiation-hydrodynamic code \cite{michel2009tuning,igumenshchev2010crossed,marion2016modeling,colaitis2021inverse,follett2022validation} due to the agreement between linear theory and experiments \cite{turnbull2020impact,hansen2021cross}. Despite this, it is still one of the most intensive procedures, increasing 3D radiation-hydrodynamic simulation expense by about $\times 5$ and limiting the number that can be run for optimization.

The repointing of the beams towards the equator, often used in PDD, leads to a changing distribution of energy absorption over time due to the expanding plasma and increasing prevalence of CBET. At early times, the plasma has not had time to expand and laser energy is deposited near the critical surface without significant CBET. This period is important as the target is most susceptible to imprint of drive asymmetries from the laser. If the laser is maintained, steady state ablation occurs between the critical surface and the ablation front, which helps to smooth drive asymmetries. Inverse bremsstrahlung deposits energy along the refracting beam's path through the plasma. Beams travelling directly up the density gradient deposit energy at higher densities, driving the target more efficiently \cite{manheimer1982steady,scheiner2019role}. This geometric effect is then exacerbated by CBET which transfers energy primarily from high energy incoming beams to refracted outgoing beams. Both these effects reduce drive, especially at the equator. The propagation of obliquely incident laser beams through a time-varying plasma leads to time dependent drive uniformity, hence the methods used for optimizing conventional DD \cite{murakami1995irradiation,murakami2010optimization,shvydky2022optimization} will not adequately optimize PDD. To mitigate the time dependent drive uniformity, some PDD configurations optimize by iteratively changing beam inputs for CBET coupled multidimensional radiation-hydrodynamic simulations \cite{hohenberger2015polar,collins2018mitigation,marozas2018wavelength,zylstra2020enhanced}. Several of these PDD configurations have been tested on solid targets at the NIF and so can provide a benchmark.

The method presented in this Letter is an algorithmic approach for creating PDD illuminations at the NIF. The optimization method uses similar tools to previous attempts including state-of-the-art inverse ray-tracing with the effects of CBET (Ifriit) \cite{colaitis2019adaptive}, however several key approximations are made which enable new configurations to be tested without running expensive radiation-hydrodynamic codes for each iteration, this leads to of order $\times 1000$ reduction in computational expense. In addition, the optimization of inputs is automated via a numerical method, not a human expert. The whole process, requires two 3D radiation-hydrodynamic simulations, an initial simulation to generate the plasma conditions and a final simulation to test the outcome of the optimization. In this Letter, these simulations are performed using the coupled ASTER-Ifriit code \cite{igumenshchev2016three,colaitis2021inverse} which is one of several state-of-the-art codes capable of reproducing key experimental features \cite{colaitis2022prl}. Beyond the methodology, 3D simulations indicate that the configuration itself results in improved drive and convergence symmetry over the comparison, indicating its applicability for future experiments.

The NIF has 192 laser beams arranged into groups of 4, called ``quads", as shown in Figure \ref{fig:mollweide_pointings}a. Each quad enters through a different port on the target chamber, and they have independent beam pointing and power balance \cite{miller2004national,keane2014national}. The quads are arranged into groups at equal angle from the poles: $\theta_p = 23.5^\circ $, $30.0^\circ $, $44.5^\circ $, and $50.0^\circ$ which are described as ``cones". The top and bottom hemisphere of the chamber are symmetric, with 4 cones in each. Each pair of cones (one in the top and bottom) have the same laser spot, however the shape and size is different between pairs. There are other parameters such as quad splitting, wavelength detuning and time varying power balance which were kept at fixed values for this optimization\cite{marozas2018first,collins2018mitigation,marozas2018wavelength,zylstra2020enhanced}.

N190204-003 is a solid target NIF experiment designed to study energy coupling for DD at megajoule scales \cite{ceurvorst2022development,viala2024cbetsolid} and is used as a benchmark in this letter. N190204-003 uses a PDD illumination, shown in Figure \ref{fig:mollweide_pointings}a, designed accounting for the impact of CBET \cite{hohenberger2015polar}. The target was $1000 \mathrm{\mu m}$ radius of deuterated plastic (CD at $1.08g/cm^3$) surrounded by $100 \mathrm{\mu m}$ of plastic (CH at $1.05g/cm^3$). The laser pulse was $4.5 \mathrm{ns}$ in total, with a two stage ramp up to a peak power of $156 \mathrm{TW}$ at $3.5 \mathrm{ns}$ \cite{viala2024cbetsolid}. Gated x-ray images \cite{kyrala2010gxd} of shock ingress were taken between, $6-8\mathrm{ns}$ with peak pressure predicted to occur at $11.7 \mathrm{ns}$ \cite{viala2024cbetsolid}.

\begin{figure}
\centering
\begin{subfigure}[t]{0.173\textwidth}
  \centering
  \topinset{(a)}{\includegraphics[width=1.0\linewidth,trim=0mm 0mm 0mm 0mm,clip]{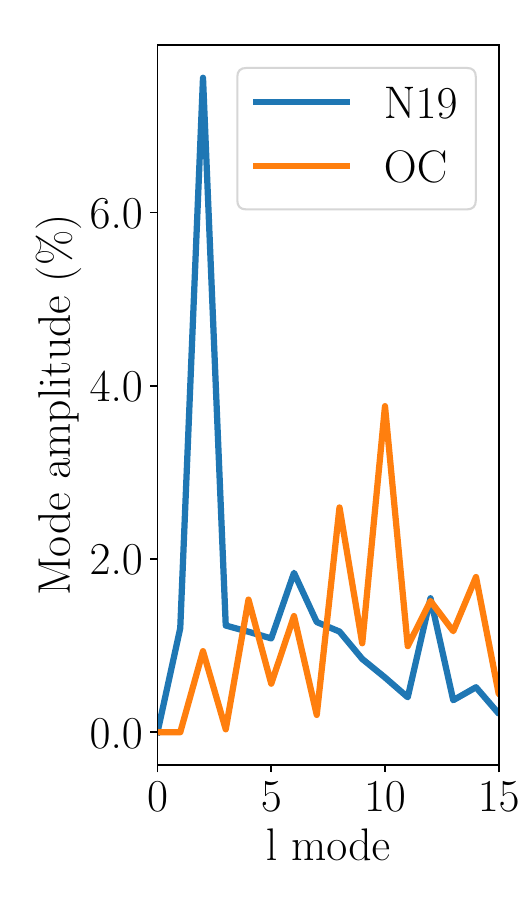}}{0pt}{0pt}
\end{subfigure}
\centering
\begin{subfigure}[t]{.15\textwidth}
  \centering
  \topinset{(b)}{\includegraphics[width=1.0\linewidth,trim=12mm 0mm 0mm 0mm,clip]{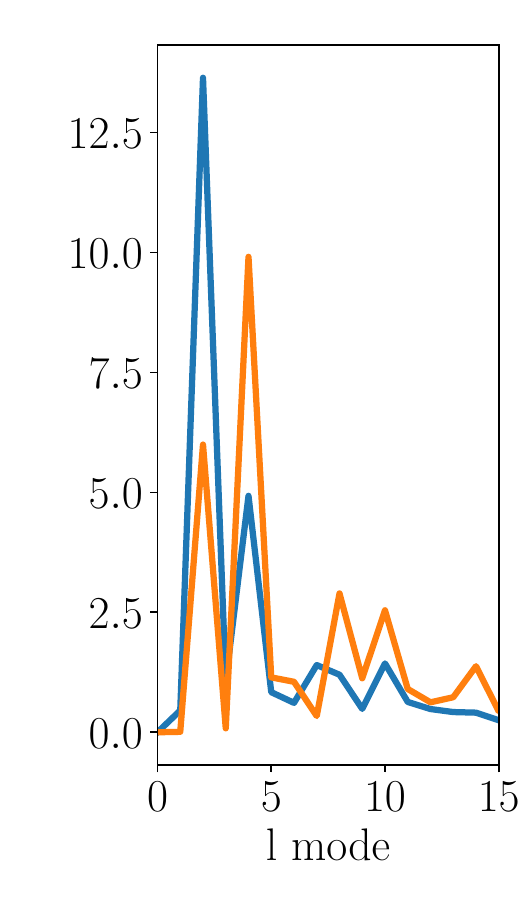}}{0pt}{0pt}
\end{subfigure}
\centering
\begin{subfigure}[t]{.15\textwidth}
  \centering
  \topinset{(c)}{\includegraphics[width=1.0\linewidth,trim=12mm 0mm 0mm 0mm,clip]{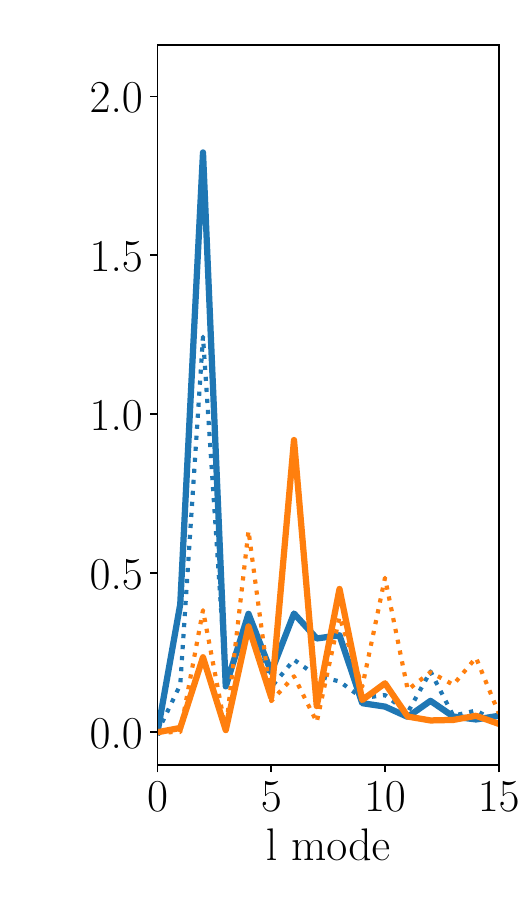}}{0pt}{0pt}
\end{subfigure}
  \caption{N190204-003 (N19, blue) and Optimized Config (OC, orange) spherical harmonics ``l modes" for: (a) illuminations without an ablation plasma, related to Figure \ref{fig:mollweide_pointings}b; (b) ablation pressures from Equation \ref{eqn:densityofabsorption} generated with plasma and CBET, related to Figure \ref{fig:mollweide_pointings}c; and (c) target areal densities at $\approx 9 \mathrm{ns}$. Mode amplitudes are given as a percentage of the mean. The dotted lines in (c) are a combination of (a) and (b) in quadrature.}
  \label{fig:modes_lineout}
\end{figure}

Several approximations were chosen to create a fast and efficient method for the evaluation of laser configurations/illuminations. Each ansatz is discussed below but validated by the overall success of the method. The approximations are: (1) a configuration can be evaluated using snapshots of the plasma at different times, (2) angularly uniform plasma profiles can be used to evaluate 3D beam configurations, (3) uniform pressure at the ablation front leads to uniform drive, (4) plasma conditions from simulating one configuration can be used to evaluate another. The first approximation is typical of finite difference methods, however there is a trade-off between time resolution and accuracy. The second and third assumptions are based on the  spherical symmetry of the plasma. The fourth approximation requires that the plasma conditions for each laser configuration evolve similarly, which is acceptable since it is only the most uniform illuminations that are of interest. 

Laser energy deposited at a lower density results in lower $P_{abl}$, ablation pressure \cite{manheimer1982steady}. Here $P_{abl}$ is generated from a weighted radial sum of the absorbed intensity,
\begin{subequations} \label{eqn:densityofabsorption}
\begin{equation} \label{eqn:manheimer}
  P_{abl} = 24.7 \ \rm{Mbar} \left( \frac{F(I_{r})}{10^{14} [\mathrm{W/cm^2}]} \right)^{2/3},
\end{equation}
\begin{equation} \label{eqn:weighting_intensity}
  F(I_{r}) = \frac{1}{R_{351}^2 n_{351}^{2/3}}\sum_{r=0}^{\infty} r^2 n_r^{2/3} I_{r},
\end{equation}
\end{subequations} 
where $R_{351}$ and $n_{351}$ are the critical radius and electron number density for wavelength $\lambda = 351 \mathrm{nm}$ and $I_{r}$ is the laser intensity absorbed ($\mathrm{W/cm^2}$) at radius $r$ with electron number density $n_r$. Equation \ref{eqn:manheimer} is based on Ref. \onlinecite{manheimer1982steady}. The $r^2$ in Equation \ref{eqn:weighting_intensity} is applied to convert intensity to units $\mathrm{W/sr}$, so it can be summed over multiple surfaces at different radii. $n_r^{2/3}$ is an empirical weighting chosen to match pressure and asymmetries observed in several 3D radiation-hydrodynamic simulations. $P_{abl}$ is calculated for all angular directions to produce an ablation pressure map, as seen in Figure \ref{fig:mollweide_pointings}c. 

In order to optimize the beam configurations, we must define a fitness function to provide a metric for comparison. Here we propose:
\begin{equation} \label{eqn:fitness_function}
    f = 10 \ \exp \left( - \left( \frac{\sigma_1^2}{9} + \frac{\sigma_2^2}{18} \right) ^{1/2} \right) \times \left( \frac{\langle P_{abl} \rangle} {50\text{Mbar}} \right)^2,
\end{equation}
where $\sigma_1$ is the standard deviation of target surface intensity for an illumination with no plasma (Figures  \ref{fig:mollweide_pointings}b and \ref{fig:modes_lineout}a)
and $\sigma_2$ is the standard deviation of the ablation pressure described in Equation \ref{eqn:densityofabsorption} (Figures \ref{fig:mollweide_pointings}c and \ref{fig:modes_lineout}b), both as percentages of the mean and only including perturbations up to spherical harmonic $l = 30$. $\langle P_{abl} \rangle$ is the angular mean of the ablation pressure from Equation \ref{eqn:densityofabsorption}. $\sigma_2$ and $\langle P_{abl} \rangle$ are evaluated in an angular averaged plasma, which is a snapshot at $4.0 \mathrm{ns}$ from a 3D radiation-hydrodynamic simulation of N190204-003. When optimizing, the same plasma is used for all evaluations, but the configuration of lasers is changed to maximize, $f(\sigma_i, \langle P_{abl} \rangle)$, in doing so ablation pressure is increased while reducing deviations from uniform illumination. Equation \ref{eqn:fitness_function} is not unique and is unlikely to be an optimal fitness function, the numerical factors were adjusted empirically but initially set as a goal for each respective term such that $f \approx 1$ upon successful optimization.

The NIF has thousands of parameters specifying each configuration, so inherent symmetries are used to reduce this number to 16. Parameters are held constant within a cone, and its symmetric pair. For each of the 4 cones within a hemisphere, 4 parameters are varied: power $[p]$, defocus $[d]$, and 2D target surface pointing $[r_s, \varphi_s]$. The power balance is bounded $2 < p < 4 \mathrm{TW/quad}$. The pointings and defocus are described relative to the location on the target surface closest to the chamber port of origin for the respective quad $[\theta_p, \phi_p]$. The location of best focus is varied along the direction of laser propagation using defocus, $0 < d < 10 \mathrm{mm}$ with $d=0.0\mathrm{mm}$ being on the target surface. Defocusing results in varying the spot size on target. Independently, the pointings are varied to cover the half of the target visible from the port. The 4 parameters are varied between the 4 cones, giving a total of 16 independent parameters.

\begin{figure}
\centering
\begin{subfigure}[t]{0.20\textwidth}
  \centering
  \topinset{(a)}{\includegraphics[width=0.7\linewidth,trim=0mm -25mm 0mm 0mm,clip]{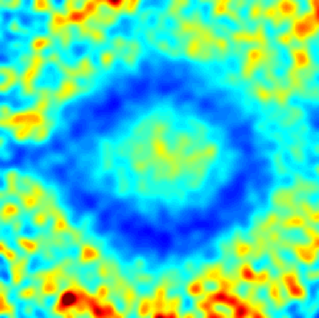}}{0pt}{-12pt}
\end{subfigure}
\centering
\begin{subfigure}[t]{0.25\textwidth}
  \centering
  \topinset{(b)}{\includegraphics[width=1.0\linewidth,trim=0mm 0mm 0mm 0mm,clip]{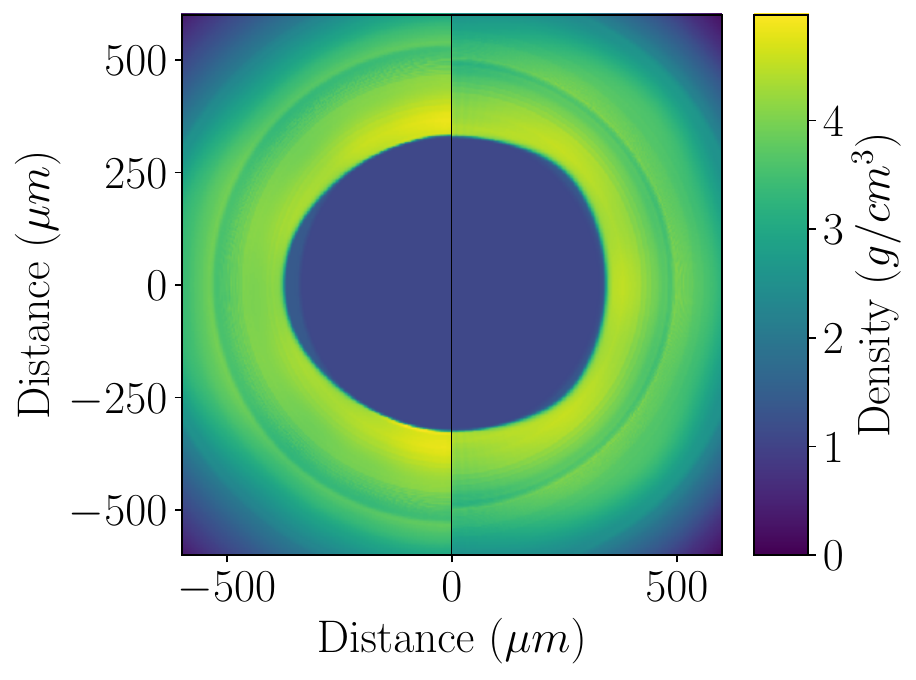}}{0pt}{0pt}
\end{subfigure}
  \caption{(a) A gated x-ray image of N190204-003 experiment at $7.73 \mathrm{ns}$. The elliptical shape of the ingoing shock is visible as the dark blue region. (b) Density slice from 3D simulation of N190204-003 at $9.00 \mathrm{ns}$ (left) and OC at $9.20 \mathrm{ns}$ (right).}
  \label{fig:2dshock_shape}
\end{figure}

\begin{table}
\begin{tabular}{|c| c c c c|} 
 \hline
 Cone & Polar angle ($^\circ$) & Offset angle ($^\circ$) & Defocus (mm) & Power (\%) \\
 \hline
 1 & 6.36 & 52.02 & 8.22 & 66.26 \\ 
 2 & 14.95 & 160.75 & 9.07 & 66.26 \\
 3 & 52.48 & -12.14 & 4.67 & 95.05 \\
 4 & 85.62 & 6.90 & 6.83 & 82.88 \\
 \hline
\end{tabular}
\vspace{-5pt}
\caption{OC parameters for each of the 4 cones in a hemisphere. The polar angle is the pointing angle from the pole, offset angle is an azimuthal offset from the port location, defocus changes the best focus along the beam propagation direction and power balance is quoted as a percent of $4 \mathrm{TW/quad}$.}
\label{tab:optimized_parameters}
\end{table}

\begin{table*}
\begin{center}
\begin{tabular}{| c | c | c | c | c | c | c | c | c | c | c | c | c | c | c | c |}
 \hline
 &&&&&\multicolumn{4}{c|}{ @$ t_3 $} & \multicolumn{4}{c|}{ @$ t_4 $} & & & \\
 \cline{6-13}
  Configuration & $\sigma_1$ & $\sigma_2$ & $\langle P_{abl} \rangle$ & $f$ & $t_3$ & $\sigma_3$ & $p_3$ & $\rho_3$ & $t_4$ & $\sigma_4$ & $p_4$ & $\rho_4$ & $N_{beams}$ & $E_{li}$ & $f_{abs}$ \\
 \hline
 N190204-003 & $8.56\%$ & $14.83\%$ & $63.5 \mathrm{Mbar}$ & $0.18$ & $9.00\mathrm{ns}$ & $2.01\%$ & $95 \mathrm{Mbars}$ & $4 \mathrm{gcc}$ & $11.7 \mathrm{ns}$ & $16.35\%$ & $4 \mathrm{Gbar}$ & $25 \mathrm{gcc}$ & 184 & $397\mathrm{kJ}$ & 82\% \\
 Optimized Config & $6.03\%$ & $12.56\%$ & $56.0 \mathrm{Mbar}$ & $0.35$ & $9.20\mathrm{ns}$ & $1.13\%$ & $80 \mathrm{Mbars}$ & $4 \mathrm{gcc}$ & $12.2 \mathrm{ns}$ & $5.56\%$ & $10 \mathrm{Gbar}$ & $35 \mathrm{gcc}$ & 192 & $408\mathrm{kJ}$ & 77\% \\
 \hline
\end{tabular}
\end{center}
\vspace{-15pt}
\caption{\label{tab:quantitive_comparison} List of simulation diagnostics values. $\sigma_1$, $\sigma_2$ and $\langle P_{abl} \rangle$ are the variables in Equation \ref{eqn:fitness_function} used to define the fitness function $f$. Time $t_3$ is after the end of the laser pulse, during shock ingress, and is chosen to match the shock radius in each simulation. At time $t_3$, $\sigma_3$ is the standard deviation in target areal density, $p_3$ is the pressure and $\rho_3$ is the density both at the shock front. The same parameters are given at $t_4$ which is the time of peak pressure in each simulation. $N_{beams}$ is the number of beams used in the simulation, $E_{li}$ is the incident laser energy and $f_{abs}$ is the absorbed laser energy percentage.}
\end{table*}

A genetic algorithm \cite{gad2021pygad}, and a coordinate descent/ascent method \cite{wright2015coordinate} were used to maximize Equation \ref{eqn:fitness_function} by varying the 16 input parameters. The methods are derivative-free meaning that exact local gradients are not available. The resultant configuration, referred to as the ``Optimized Config" (OC), does not represent a maximum in the search space, however it exceeded the fitness, according to Equation \ref{eqn:fitness_function}, of the PDD configuration used in N190204-003 within a certain allocation of compute time ($< 2 \mathrm{MCPU \ hours}$). The parameters for the OC are given in Table \ref{tab:optimized_parameters}, where the angles are given as target surface/chamber coordinates $[\theta, \phi]$. A quantitative comparison of N190204-003 and the OC is displayed in Table \ref{tab:quantitive_comparison}. The OC was found within 550 searched configurations, which consisted of 185 genetic algorithm evaluations and 365 coordinate descent evaluations. It can take $>10,000$ evaluations for the method to settle at a (local/global) maximum, determined by the optimizations of $\sigma_1$ alone. This Letter does not focus on the search procedure, which is an active area of research within mathematics/computer science, but the overall method used to evaluate illumination configurations.

Figure \ref{fig:mollweide_pointings}a shows the difference in pointing between N190204-003, and the OC. N190204-003 features ``quad-splitting" where the pointings for the 4 beams within a single quad differ, and time varying power balances, where each cone's fraction of the total laser energy varies over time, whereas the OC does not. The OC does have independent defocus for each of the cones, whereas $d=10 \mathrm{mm}$ is used for all cones in N190204-003. The overall complexity of each configuration is similar, and it is likely that they each present a similar level of challenge to recreate experimentally. Figure \ref{fig:mollweide_pointings}a shows some of the beams in N190204-003 that were perturbed to account for the target stalk, in addition, two of the quads were used for an x-ray backlighter. These modifications had only a minor effect on the overall illumination uniformity (improving $\sigma_i$ by $<1\%$). The azimuthal offset pointings for OC were selected to go the same direction in each hemisphere, despite this not conforming to the rotational symmetry of the ports. This was done to reduce CBET from beams travelling in opposite directions at the equator. The large offset angles given for cones 1 and 2 of OC in Table \ref{tab:optimized_parameters} indicate that offsetting the most polar cones to create more oblique incidence angles can improve uniformity by mimicking the angles required to get energy to the equator.

Figure \ref{fig:mollweide_pointings}b shows the two configurations illuminating a $1100 \mathrm{\mu m}$ radius target without a plasma, this is similar to the absorption that occurs at the start of the laser pulse ($<0.5 \mathrm{ns}$). The spherical modes for the two illuminations are given in Figure \ref{fig:modes_lineout}a and $\sigma_1$ is given in Table \ref{tab:quantitive_comparison}. N190204-003 has a larger mode 2 which leads to it having a larger overall $\sigma_1$. Figure \ref{fig:mollweide_pointings}c is created by illuminating the angularly averaged plasma conditions from a 3D radiation-hydrodynamic simulation of N190204-003 at $4\mathrm{ns}$. The OC has a smaller mode 2 which can also be seen in Figure \ref{fig:modes_lineout}b, but a larger mode 4. Table \ref{tab:quantitive_comparison} shows that the OC has a smaller overall $\sigma_2$ but also a predicted reduction in $\langle P_{abl} \rangle$ when compared to N190204-003.

Once an illumination configuration is selected for maximizing Equation \ref{eqn:fitness_function}, it is then tested with a full 3D radiation-hydrodynamics simulation. The cone power values given in Table \ref{tab:optimized_parameters} have been rescaled so that $E_{li}$ matches the experimentally requested laser pulse of N190204-003. The rescaling increased drive asymmetry of OC from $\sigma_2 = 8.83 \%$ due to the non-linear effects of CBET. N190204-003 uses the delivered pulse while OC is simulated with the requested pulse, leading to the difference in incident energy, $E_{li}$ in Table \ref{tab:quantitive_comparison}. Despite this, N190204-003 couples more energy to the target. 

Figures \ref{fig:modes_lineout}c shows the modes in areal density for simulation of the N190204-003 configuration at $9.0\mathrm{ns}$ and the OC at $9.2 \mathrm{ns}$. The times are chosen so that the shock radius matches, as can be seen in Figure \ref{fig:2dshock_shape} (b). Figures \ref{fig:modes_lineout}c also shows dotted lines, which are a combination of the modes from Figures \ref{fig:modes_lineout}a and \ref{fig:modes_lineout}b in quadrature, with the same weightings used in Equation \ref{eqn:fitness_function}. It shows the strengths and weaknesses of the initial approximations. The inaccuracy predicting the higher modes is likely due to the Bell-Plesset effect \cite{epstein2004bell} while the OC also features a mode 6 in the areal density, which is not seen in the two snapshots, this is possibly caused by inhomogeneous plasma effects, or lacking temporal resolution. The matching of modes is not a requirement for the method to work, however the similarity is evidence that reinforces the initial assumptions.

Figure \ref{fig:2dshock_shape}a shows a flat-field corrected, gated x-ray image \cite{kyrala2010gxd} of the N190204-003 shock at $7.73 \mathrm{ns}$. The simulation of N190204-003 shown in Figure \ref{fig:2dshock_shape}b has a similar shape but at a later time, $9.0\mathrm{ns}$. Figure \ref{fig:2dshock_shape}b shows the improved shape that is achieved when using the OC for illuminating the target. This is reinforced by comparing $\sigma_3$ in Table \ref{tab:quantitive_comparison}. N190204-003 features an earlier peak pressure at $t_4 = 11.7 \mathrm{ns}$ but $\sigma_4$ is approximately $\times 3$ that of the OC, resulting in N190204-003's lower $p_4$ and $\rho_4$ despite coupling more energy to the target. In this deuterated solid target experiment, higher density and pressure at shock convergence could create the conditions necessary for fusion, giving an x-ray flash which is a useful diagnostic but was not observed in N190204-003.

The simulated drive uniformity demonstrated in this paper is an important step, however, more is required to use PDD to drive an implosion to ignition at the NIF. It has not been demonstrated that both $\sigma_1$ and $\sigma_2$ can be reduced below $\approx2\%$ required for a high performance implosion. Using a similar method, it may be possible to achieve the requisite $\sigma_i<2\%$ for all times if the power balance was varied between the snapshots. In addition, it is likely that for implosions, more snapshots will need to be considered. Time varying power balance would add a new parameter per cone per snapshot, and so principal component analysis \cite{bro2014principal} could be used to reduce the parameters space.

Demonstrated in this letter is a new, efficient, algorithmic approach to creating illumination configurations for PDD solid targets. This alone is a critical development, as previously the method for development have required many hours from a human expert, alongside numerous radiation-hydrodynamic simulations. Beyond this the illumination configuration is novel, practical and results in higher peak pressure and density when simulated by 3D radiation-hydrodynamics. The method can be modified for implosion targets and the process is not limited to spherical PDD, but could provide efficient optimization where illumination uniformity is important, regardless of geometry, including hohlraums. It could also be vital for DD as future designs are expected to require many beams per port to balance drive uniformity \cite{colaitis2023exploration,eimerl2014stardriver}, against chamber efficiency. DD with multiple beams per port can no longer rely on traditional techniques and will require iterative optimization such as the method presented here to achieve the necessary drive uniformity.

\begin{acknowledgments}
We would like to acknowledge support and funding from the CNRS, CEA, and Universit\'{e} de Bordeaux including the staff of CELIA. This work has been carried out within the framework of the EUROfusion Consortium, funded by the European Union via the Euratom Research and Training Programme (Grant Agreement No. 101052200–EUROfusion). Views and opinions expressed are, however, those of the author(s) only and do not necessarily reflect those of the European Union or the European Commission. Neither the European Union nor the European Commission can be held responsible for them. The involved teams have operated within the framework of the Enabling Research Project: ENR-IFE.01.CEA ‘Advancing shock ignition for direct-drive inertial fusion’. This work was granted access to the HPC resources of TGCC under the Allocation Nos. 2023-A0140514117 made by GENCI. This material is based on work supported by the Department of Energy National Nuclear Security Administration under Award No. DE-NA0003856, the University of Rochester, and the New York State Energy Research and Development Authority. M.J.V.S. acknowledges support from the Royal Society URF-R1221874.

The code and data used for optimization are available via an MIT license from Github: https://github.com/DuncanBarlow/PDDOptimisation, release tag: v0.1.0-alpha. For access to the other codes used please contact the author. The data will also be uploaded to a repository linked to the journal.

\end{acknowledgments}

\bibliography{aipsamp}

\providecommand{\noopsort}[1]{}\providecommand{\singleletter}[1]{#1}%
\begin{thebibliography}{50}%
\makeatletter
\providecommand \@ifxundefined [1]{%
 \@ifx{#1\undefined}
}%
\providecommand \@ifnum [1]{%
 \ifnum #1\expandafter \@firstoftwo
 \else \expandafter \@secondoftwo
 \fi
}%
\providecommand \@ifx [1]{%
 \ifx #1\expandafter \@firstoftwo
 \else \expandafter \@secondoftwo
 \fi
}%
\providecommand \natexlab [1]{#1}%
\providecommand \enquote  [1]{``#1''}%
\providecommand \bibnamefont  [1]{#1}%
\providecommand \bibfnamefont [1]{#1}%
\providecommand \citenamefont [1]{#1}%
\providecommand \href@noop [0]{\@secondoftwo}%
\providecommand \href [0]{\begingroup \@sanitize@url \@href}%
\providecommand \@href[1]{\@@startlink{#1}\@@href}%
\providecommand \@@href[1]{\endgroup#1\@@endlink}%
\providecommand \@sanitize@url [0]{\catcode `\\12\catcode `\$12\catcode
  `\&12\catcode `\#12\catcode `\^12\catcode `\_12\catcode `\%12\relax}%
\providecommand \@@startlink[1]{}%
\providecommand \@@endlink[0]{}%
\providecommand \url  [0]{\begingroup\@sanitize@url \@url }%
\providecommand \@url [1]{\endgroup\@href {#1}{\urlprefix }}%
\providecommand \urlprefix  [0]{URL }%
\providecommand \Eprint [0]{\href }%
\providecommand \doibase [0]{http://dx.doi.org/}%
\providecommand \selectlanguage [0]{\@gobble}%
\providecommand \bibinfo  [0]{\@secondoftwo}%
\providecommand \bibfield  [0]{\@secondoftwo}%
\providecommand \translation [1]{[#1]}%
\providecommand \BibitemOpen [0]{}%
\providecommand \bibitemStop [0]{}%
\providecommand \bibitemNoStop [0]{.\EOS\space}%
\providecommand \EOS [0]{\spacefactor3000\relax}%
\providecommand \BibitemShut  [1]{\csname bibitem#1\endcsname}%
\let\auto@bib@innerbib\@empty
\bibitem [{\citenamefont {Miller}, \citenamefont {Moses},\ and\ \citenamefont
  {Wuest}(2004)}]{miller2004national}%
  \BibitemOpen
  \bibfield  {author} {\bibinfo {author} {\bibfnamefont {G.~H.}\ \bibnamefont
  {Miller}}, \bibinfo {author} {\bibfnamefont {E.~I.}\ \bibnamefont {Moses}}, \
  and\ \bibinfo {author} {\bibfnamefont {C.~R.}\ \bibnamefont {Wuest}},\
  }\bibfield  {title} {\enquote {\bibinfo {title} {The national ignition
  facility},}\ }\href@noop {} {\bibfield  {journal} {\bibinfo  {journal}
  {Optical Engineering}\ }\textbf {\bibinfo {volume} {43}},\ \bibinfo {pages}
  {2841--2853} (\bibinfo {year} {2004})}\BibitemShut {NoStop}%
\bibitem [{\citenamefont {Miquel}, \citenamefont {Lion},\ and\ \citenamefont
  {Vivini}(2016)}]{miquel2016laser}%
  \BibitemOpen
  \bibfield  {author} {\bibinfo {author} {\bibfnamefont {J.}~\bibnamefont
  {Miquel}}, \bibinfo {author} {\bibfnamefont {C.}~\bibnamefont {Lion}}, \ and\
  \bibinfo {author} {\bibfnamefont {P.}~\bibnamefont {Vivini}},\ }\bibfield
  {title} {\enquote {\bibinfo {title} {The laser mega-joule: Lmj \& petal
  status and program overview},}\ }in\ \href@noop {} {\emph {\bibinfo
  {booktitle} {Journal of Physics: Conference Series}}},\ Vol.\ \bibinfo
  {volume} {688}\ (\bibinfo {organization} {IOP Publishing},\ \bibinfo {year}
  {2016})\ p.\ \bibinfo {pages} {012067}\BibitemShut {NoStop}%
\bibitem [{\citenamefont {Nuckolls}\ \emph {et~al.}(1972)\citenamefont
  {Nuckolls}, \citenamefont {Wood}, \citenamefont {Thiessen},\ and\
  \citenamefont {Zimmerman}}]{nuckolls1972laser}%
  \BibitemOpen
  \bibfield  {author} {\bibinfo {author} {\bibfnamefont {J.}~\bibnamefont
  {Nuckolls}}, \bibinfo {author} {\bibfnamefont {L.}~\bibnamefont {Wood}},
  \bibinfo {author} {\bibfnamefont {A.}~\bibnamefont {Thiessen}}, \ and\
  \bibinfo {author} {\bibfnamefont {G.}~\bibnamefont {Zimmerman}},\ }\bibfield
  {title} {\enquote {\bibinfo {title} {Laser compression of matter to
  super-high densities: Thermonuclear (ctr) applications},}\ }\href@noop {}
  {\bibfield  {journal} {\bibinfo  {journal} {Nature}\ }\textbf {\bibinfo
  {volume} {239}},\ \bibinfo {pages} {139--142} (\bibinfo {year}
  {1972})}\BibitemShut {NoStop}%
\bibitem [{\citenamefont {Atzeni}\ and\ \citenamefont {Meyer-ter
  Vehn}(2004)}]{atzeni2004physics}%
  \BibitemOpen
  \bibfield  {author} {\bibinfo {author} {\bibfnamefont {S.}~\bibnamefont
  {Atzeni}}\ and\ \bibinfo {author} {\bibfnamefont {J.}~\bibnamefont {Meyer-ter
  Vehn}},\ }\href@noop {} {\emph {\bibinfo {title} {The physics of inertial
  fusion: beam plasma interaction, hydrodynamics, hot dense matter}}},\ Vol.\
  \bibinfo {volume} {125}\ (\bibinfo  {publisher} {OUP Oxford},\ \bibinfo
  {year} {2004})\BibitemShut {NoStop}%
\bibitem [{\citenamefont {Lindl}\ \emph {et~al.}(2004)\citenamefont {Lindl},
  \citenamefont {Amendt}, \citenamefont {Berger}, \citenamefont {Glendinning},
  \citenamefont {Glenzer}, \citenamefont {Haan}, \citenamefont {Kauffman},
  \citenamefont {Landen},\ and\ \citenamefont {Suter}}]{lindl2004physics}%
  \BibitemOpen
  \bibfield  {author} {\bibinfo {author} {\bibfnamefont {J.~D.}\ \bibnamefont
  {Lindl}}, \bibinfo {author} {\bibfnamefont {P.}~\bibnamefont {Amendt}},
  \bibinfo {author} {\bibfnamefont {R.~L.}\ \bibnamefont {Berger}}, \bibinfo
  {author} {\bibfnamefont {S.~G.}\ \bibnamefont {Glendinning}}, \bibinfo
  {author} {\bibfnamefont {S.~H.}\ \bibnamefont {Glenzer}}, \bibinfo {author}
  {\bibfnamefont {S.~W.}\ \bibnamefont {Haan}}, \bibinfo {author}
  {\bibfnamefont {R.~L.}\ \bibnamefont {Kauffman}}, \bibinfo {author}
  {\bibfnamefont {O.~L.}\ \bibnamefont {Landen}}, \ and\ \bibinfo {author}
  {\bibfnamefont {L.~J.}\ \bibnamefont {Suter}},\ }\bibfield  {title} {\enquote
  {\bibinfo {title} {The physics basis for ignition using indirect-drive
  targets on the national ignition facility},}\ }\href@noop {} {\bibfield
  {journal} {\bibinfo  {journal} {Physics of plasmas}\ }\textbf {\bibinfo
  {volume} {11}},\ \bibinfo {pages} {339--491} (\bibinfo {year}
  {2004})}\BibitemShut {NoStop}%
\bibitem [{\citenamefont {Li}\ \emph {et~al.}(2004)\citenamefont {Li},
  \citenamefont {S{\'e}guin}, \citenamefont {Frenje}, \citenamefont {Petrasso},
  \citenamefont {Delettrez}, \citenamefont {McKenty}, \citenamefont {Sangster},
  \citenamefont {Keck}, \citenamefont {Soures}, \citenamefont {Marshall} \emph
  {et~al.}}]{li2004effects}%
  \BibitemOpen
  \bibfield  {author} {\bibinfo {author} {\bibfnamefont {C.}~\bibnamefont
  {Li}}, \bibinfo {author} {\bibfnamefont {F.}~\bibnamefont {S{\'e}guin}},
  \bibinfo {author} {\bibfnamefont {J.}~\bibnamefont {Frenje}}, \bibinfo
  {author} {\bibfnamefont {R.}~\bibnamefont {Petrasso}}, \bibinfo {author}
  {\bibfnamefont {J.}~\bibnamefont {Delettrez}}, \bibinfo {author}
  {\bibfnamefont {P.}~\bibnamefont {McKenty}}, \bibinfo {author} {\bibfnamefont
  {T.}~\bibnamefont {Sangster}}, \bibinfo {author} {\bibfnamefont
  {R.}~\bibnamefont {Keck}}, \bibinfo {author} {\bibfnamefont {J.}~\bibnamefont
  {Soures}}, \bibinfo {author} {\bibfnamefont {F.}~\bibnamefont {Marshall}},
  \emph {et~al.},\ }\bibfield  {title} {\enquote {\bibinfo {title} {Effects of
  nonuniform illumination on implosion asymmetry in direct-drive inertial
  confinement fusion},}\ }\href@noop {} {\bibfield  {journal} {\bibinfo
  {journal} {Physical review letters}\ }\textbf {\bibinfo {volume} {92}},\
  \bibinfo {pages} {205001} (\bibinfo {year} {2004})}\BibitemShut {NoStop}%
\bibitem [{\citenamefont {Abu-Shawareb}\ \emph {et~al.}(2022)\citenamefont
  {Abu-Shawareb}, \citenamefont {Acree}, \citenamefont {Adams}, \citenamefont
  {Adams}, \citenamefont {Addis}, \citenamefont {Aden}, \citenamefont {Adrian},
  \citenamefont {Afeyan}, \citenamefont {Aggleton}, \citenamefont {Aghaian}
  \emph {et~al.}}]{abu2022lawson}%
  \BibitemOpen
  \bibfield  {author} {\bibinfo {author} {\bibfnamefont {H.}~\bibnamefont
  {Abu-Shawareb}}, \bibinfo {author} {\bibfnamefont {R.}~\bibnamefont {Acree}},
  \bibinfo {author} {\bibfnamefont {P.}~\bibnamefont {Adams}}, \bibinfo
  {author} {\bibfnamefont {J.}~\bibnamefont {Adams}}, \bibinfo {author}
  {\bibfnamefont {B.}~\bibnamefont {Addis}}, \bibinfo {author} {\bibfnamefont
  {R.}~\bibnamefont {Aden}}, \bibinfo {author} {\bibfnamefont {P.}~\bibnamefont
  {Adrian}}, \bibinfo {author} {\bibfnamefont {B.}~\bibnamefont {Afeyan}},
  \bibinfo {author} {\bibfnamefont {M.}~\bibnamefont {Aggleton}}, \bibinfo
  {author} {\bibfnamefont {L.}~\bibnamefont {Aghaian}},  \emph {et~al.},\
  }\bibfield  {title} {\enquote {\bibinfo {title} {Lawson criterion for
  ignition exceeded in an inertial fusion experiment},}\ }\href@noop {}
  {\bibfield  {journal} {\bibinfo  {journal} {Physical review letters}\
  }\textbf {\bibinfo {volume} {129}},\ \bibinfo {pages} {075001} (\bibinfo
  {year} {2022})}\BibitemShut {NoStop}%
\bibitem [{\citenamefont {Craxton}\ \emph {et~al.}(2015)\citenamefont
  {Craxton}, \citenamefont {Anderson}, \citenamefont {Boehly}, \citenamefont
  {Goncharov}, \citenamefont {Harding}, \citenamefont {Knauer}, \citenamefont
  {McCrory}, \citenamefont {McKenty}, \citenamefont {Meyerhofer}, \citenamefont
  {Myatt} \emph {et~al.}}]{craxton2015direct}%
  \BibitemOpen
  \bibfield  {author} {\bibinfo {author} {\bibfnamefont {R.}~\bibnamefont
  {Craxton}}, \bibinfo {author} {\bibfnamefont {K.}~\bibnamefont {Anderson}},
  \bibinfo {author} {\bibfnamefont {T.}~\bibnamefont {Boehly}}, \bibinfo
  {author} {\bibfnamefont {V.}~\bibnamefont {Goncharov}}, \bibinfo {author}
  {\bibfnamefont {D.}~\bibnamefont {Harding}}, \bibinfo {author} {\bibfnamefont
  {J.}~\bibnamefont {Knauer}}, \bibinfo {author} {\bibfnamefont
  {R.}~\bibnamefont {McCrory}}, \bibinfo {author} {\bibfnamefont
  {P.}~\bibnamefont {McKenty}}, \bibinfo {author} {\bibfnamefont
  {D.}~\bibnamefont {Meyerhofer}}, \bibinfo {author} {\bibfnamefont
  {J.}~\bibnamefont {Myatt}},  \emph {et~al.},\ }\bibfield  {title} {\enquote
  {\bibinfo {title} {Direct-drive inertial confinement fusion: A review},}\
  }\href@noop {} {\bibfield  {journal} {\bibinfo  {journal} {Physics of
  Plasmas}\ }\textbf {\bibinfo {volume} {22}} (\bibinfo {year}
  {2015})}\BibitemShut {NoStop}%
\bibitem [{\citenamefont {Cola{\"\i}tis}\ \emph {et~al.}(2022)\citenamefont
  {Cola{\"\i}tis}, \citenamefont {Turnbull}, \citenamefont {Igumenschev},
  \citenamefont {Edgell}, \citenamefont {Shah}, \citenamefont {Mannion},
  \citenamefont {Stoeckl}, \citenamefont {Jacob-Perkins}, \citenamefont
  {Shvydky}, \citenamefont {Janezic} \emph {et~al.}}]{colaitis2022prl}%
  \BibitemOpen
  \bibfield  {author} {\bibinfo {author} {\bibfnamefont {A.}~\bibnamefont
  {Cola{\"\i}tis}}, \bibinfo {author} {\bibfnamefont {D.}~\bibnamefont
  {Turnbull}}, \bibinfo {author} {\bibfnamefont {I.}~\bibnamefont
  {Igumenschev}}, \bibinfo {author} {\bibfnamefont {D.}~\bibnamefont {Edgell}},
  \bibinfo {author} {\bibfnamefont {R.}~\bibnamefont {Shah}}, \bibinfo {author}
  {\bibfnamefont {O.}~\bibnamefont {Mannion}}, \bibinfo {author} {\bibfnamefont
  {C.}~\bibnamefont {Stoeckl}}, \bibinfo {author} {\bibfnamefont
  {D.}~\bibnamefont {Jacob-Perkins}}, \bibinfo {author} {\bibfnamefont
  {A.}~\bibnamefont {Shvydky}}, \bibinfo {author} {\bibfnamefont
  {R.}~\bibnamefont {Janezic}},  \emph {et~al.},\ }\bibfield  {title} {\enquote
  {\bibinfo {title} {3d simulations capture the persistent low-mode asymmetries
  evident in laser-direct-drive implosions on omega},}\ }\href@noop {}
  {\bibfield  {journal} {\bibinfo  {journal} {Physical Review Letters}\
  }\textbf {\bibinfo {volume} {129}},\ \bibinfo {pages} {095001} (\bibinfo
  {year} {2022})}\BibitemShut {NoStop}%
\bibitem [{\citenamefont {Boehly}\ \emph {et~al.}(1997)\citenamefont {Boehly},
  \citenamefont {Brown}, \citenamefont {Craxton}, \citenamefont {Keck},
  \citenamefont {Knauer}, \citenamefont {Kelly}, \citenamefont {Kessler},
  \citenamefont {Kumpan}, \citenamefont {Loucks}, \citenamefont {Letzring}
  \emph {et~al.}}]{boehly1997omega}%
  \BibitemOpen
  \bibfield  {author} {\bibinfo {author} {\bibfnamefont {T.}~\bibnamefont
  {Boehly}}, \bibinfo {author} {\bibfnamefont {D.}~\bibnamefont {Brown}},
  \bibinfo {author} {\bibfnamefont {R.}~\bibnamefont {Craxton}}, \bibinfo
  {author} {\bibfnamefont {R.}~\bibnamefont {Keck}}, \bibinfo {author}
  {\bibfnamefont {J.}~\bibnamefont {Knauer}}, \bibinfo {author} {\bibfnamefont
  {J.}~\bibnamefont {Kelly}}, \bibinfo {author} {\bibfnamefont
  {T.}~\bibnamefont {Kessler}}, \bibinfo {author} {\bibfnamefont
  {S.}~\bibnamefont {Kumpan}}, \bibinfo {author} {\bibfnamefont
  {S.}~\bibnamefont {Loucks}}, \bibinfo {author} {\bibfnamefont
  {S.}~\bibnamefont {Letzring}},  \emph {et~al.},\ }\bibfield  {title}
  {\enquote {\bibinfo {title} {Initial performance results of the omega laser
  system},}\ }\href@noop {} {\bibfield  {journal} {\bibinfo  {journal} {Optics
  communications}\ }\textbf {\bibinfo {volume} {133}},\ \bibinfo {pages}
  {495--506} (\bibinfo {year} {1997})}\BibitemShut {NoStop}%
\bibitem [{\citenamefont {Nora}\ \emph {et~al.}(2014)\citenamefont {Nora},
  \citenamefont {Betti}, \citenamefont {Anderson}, \citenamefont {Shvydky},
  \citenamefont {Bose}, \citenamefont {Woo}, \citenamefont {Christopherson},
  \citenamefont {Marozas}, \citenamefont {Collins}, \citenamefont {Radha} \emph
  {et~al.}}]{nora2014theory}%
  \BibitemOpen
  \bibfield  {author} {\bibinfo {author} {\bibfnamefont {R.}~\bibnamefont
  {Nora}}, \bibinfo {author} {\bibfnamefont {R.}~\bibnamefont {Betti}},
  \bibinfo {author} {\bibfnamefont {K.}~\bibnamefont {Anderson}}, \bibinfo
  {author} {\bibfnamefont {A.}~\bibnamefont {Shvydky}}, \bibinfo {author}
  {\bibfnamefont {A.}~\bibnamefont {Bose}}, \bibinfo {author} {\bibfnamefont
  {K.}~\bibnamefont {Woo}}, \bibinfo {author} {\bibfnamefont {A.}~\bibnamefont
  {Christopherson}}, \bibinfo {author} {\bibfnamefont {J.}~\bibnamefont
  {Marozas}}, \bibinfo {author} {\bibfnamefont {T.}~\bibnamefont {Collins}},
  \bibinfo {author} {\bibfnamefont {P.}~\bibnamefont {Radha}},  \emph
  {et~al.},\ }\bibfield  {title} {\enquote {\bibinfo {title} {Theory of
  hydro-equivalent ignition for inertial fusion and its applications to omega
  and the national ignition facility},}\ }\href@noop {} {\bibfield  {journal}
  {\bibinfo  {journal} {Physics of Plasmas}\ }\textbf {\bibinfo {volume} {21}}
  (\bibinfo {year} {2014})}\BibitemShut {NoStop}%
\bibitem [{\citenamefont {Skupsky}\ \emph {et~al.}(2004)\citenamefont
  {Skupsky}, \citenamefont {Marozas}, \citenamefont {Craxton}, \citenamefont
  {Betti}, \citenamefont {Collins}, \citenamefont {Delettrez}, \citenamefont
  {Goncharov}, \citenamefont {McKenty}, \citenamefont {Radha}, \citenamefont
  {Boehly} \emph {et~al.}}]{skupsky2004polar}%
  \BibitemOpen
  \bibfield  {author} {\bibinfo {author} {\bibfnamefont {S.}~\bibnamefont
  {Skupsky}}, \bibinfo {author} {\bibfnamefont {J.}~\bibnamefont {Marozas}},
  \bibinfo {author} {\bibfnamefont {R.}~\bibnamefont {Craxton}}, \bibinfo
  {author} {\bibfnamefont {R.}~\bibnamefont {Betti}}, \bibinfo {author}
  {\bibfnamefont {T.}~\bibnamefont {Collins}}, \bibinfo {author} {\bibfnamefont
  {J.}~\bibnamefont {Delettrez}}, \bibinfo {author} {\bibfnamefont
  {V.}~\bibnamefont {Goncharov}}, \bibinfo {author} {\bibfnamefont
  {P.}~\bibnamefont {McKenty}}, \bibinfo {author} {\bibfnamefont
  {P.}~\bibnamefont {Radha}}, \bibinfo {author} {\bibfnamefont
  {T.}~\bibnamefont {Boehly}},  \emph {et~al.},\ }\bibfield  {title} {\enquote
  {\bibinfo {title} {Polar direct drive on the national ignition facility},}\
  }\href@noop {} {\bibfield  {journal} {\bibinfo  {journal} {Physics of
  Plasmas}\ }\textbf {\bibinfo {volume} {11}},\ \bibinfo {pages} {2763--2770}
  (\bibinfo {year} {2004})}\BibitemShut {NoStop}%
\bibitem [{\citenamefont {Marozas}\ \emph {et~al.}(2006)\citenamefont
  {Marozas}, \citenamefont {Marshall}, \citenamefont {Craxton}, \citenamefont
  {Igumenshchev}, \citenamefont {Skupsky}, \citenamefont {Bonino},
  \citenamefont {Collins}, \citenamefont {Epstein}, \citenamefont {Glebov},
  \citenamefont {Jacobs-Perkins} \emph {et~al.}}]{marozas2006polar}%
  \BibitemOpen
  \bibfield  {author} {\bibinfo {author} {\bibfnamefont {J.}~\bibnamefont
  {Marozas}}, \bibinfo {author} {\bibfnamefont {F.}~\bibnamefont {Marshall}},
  \bibinfo {author} {\bibfnamefont {R.}~\bibnamefont {Craxton}}, \bibinfo
  {author} {\bibfnamefont {I.}~\bibnamefont {Igumenshchev}}, \bibinfo {author}
  {\bibfnamefont {S.}~\bibnamefont {Skupsky}}, \bibinfo {author} {\bibfnamefont
  {M.}~\bibnamefont {Bonino}}, \bibinfo {author} {\bibfnamefont
  {T.}~\bibnamefont {Collins}}, \bibinfo {author} {\bibfnamefont
  {R.}~\bibnamefont {Epstein}}, \bibinfo {author} {\bibfnamefont {V.~Y.}\
  \bibnamefont {Glebov}}, \bibinfo {author} {\bibfnamefont {D.}~\bibnamefont
  {Jacobs-Perkins}},  \emph {et~al.},\ }\bibfield  {title} {\enquote {\bibinfo
  {title} {Polar-direct-drive simulations and experiments},}\ }\href@noop {}
  {\bibfield  {journal} {\bibinfo  {journal} {Physics of plasmas}\ }\textbf
  {\bibinfo {volume} {13}} (\bibinfo {year} {2006})}\BibitemShut {NoStop}%
\bibitem [{\citenamefont {Canaud}\ \emph {et~al.}(2007)\citenamefont {Canaud},
  \citenamefont {Garaude}, \citenamefont {Clique}, \citenamefont {Lecler},
  \citenamefont {Masson}, \citenamefont {Quach},\ and\ \citenamefont {Van~der
  Vliet}}]{canaud2007high}%
  \BibitemOpen
  \bibfield  {author} {\bibinfo {author} {\bibfnamefont {B.}~\bibnamefont
  {Canaud}}, \bibinfo {author} {\bibfnamefont {F.}~\bibnamefont {Garaude}},
  \bibinfo {author} {\bibfnamefont {C.}~\bibnamefont {Clique}}, \bibinfo
  {author} {\bibfnamefont {N.}~\bibnamefont {Lecler}}, \bibinfo {author}
  {\bibfnamefont {A.}~\bibnamefont {Masson}}, \bibinfo {author} {\bibfnamefont
  {R.}~\bibnamefont {Quach}}, \ and\ \bibinfo {author} {\bibfnamefont
  {J.}~\bibnamefont {Van~der Vliet}},\ }\bibfield  {title} {\enquote {\bibinfo
  {title} {High-gain direct-drive laser fusion with indirect drive beam layout
  of laser m{\'e}gajoule},}\ }\href@noop {} {\bibfield  {journal} {\bibinfo
  {journal} {Nuclear fusion}\ }\textbf {\bibinfo {volume} {47}},\ \bibinfo
  {pages} {1652} (\bibinfo {year} {2007})}\BibitemShut {NoStop}%
\bibitem [{\citenamefont {Yeamans}\ \emph {et~al.}(2021)\citenamefont
  {Yeamans}, \citenamefont {Kemp}, \citenamefont {Walters}, \citenamefont
  {Whitley}, \citenamefont {McKenty}, \citenamefont {Garcia}, \citenamefont
  {Yang}, \citenamefont {Craxton},\ and\ \citenamefont
  {Blue}}]{yeamans2021high}%
  \BibitemOpen
  \bibfield  {author} {\bibinfo {author} {\bibfnamefont {C.}~\bibnamefont
  {Yeamans}}, \bibinfo {author} {\bibfnamefont {G.}~\bibnamefont {Kemp}},
  \bibinfo {author} {\bibfnamefont {Z.}~\bibnamefont {Walters}}, \bibinfo
  {author} {\bibfnamefont {H.}~\bibnamefont {Whitley}}, \bibinfo {author}
  {\bibfnamefont {P.}~\bibnamefont {McKenty}}, \bibinfo {author} {\bibfnamefont
  {E.}~\bibnamefont {Garcia}}, \bibinfo {author} {\bibfnamefont
  {Y.}~\bibnamefont {Yang}}, \bibinfo {author} {\bibfnamefont {R.}~\bibnamefont
  {Craxton}}, \ and\ \bibinfo {author} {\bibfnamefont {B.}~\bibnamefont
  {Blue}},\ }\bibfield  {title} {\enquote {\bibinfo {title} {High yield polar
  direct drive fusion neutron sources at the national ignition facility},}\
  }\href@noop {} {\bibfield  {journal} {\bibinfo  {journal} {Nuclear Fusion}\
  }\textbf {\bibinfo {volume} {61}},\ \bibinfo {pages} {046031} (\bibinfo
  {year} {2021})}\BibitemShut {NoStop}%
\bibitem [{\citenamefont {Regan}\ \emph {et~al.}(2018)\citenamefont {Regan},
  \citenamefont {Goncharov}, \citenamefont {Sangster}, \citenamefont
  {Campbell}, \citenamefont {Betti}, \citenamefont {Anderson}, \citenamefont
  {Bernat}, \citenamefont {Bose}, \citenamefont {Boehly}, \citenamefont
  {Bonino} \emph {et~al.}}]{regan2018national}%
  \BibitemOpen
  \bibfield  {author} {\bibinfo {author} {\bibfnamefont {S.}~\bibnamefont
  {Regan}}, \bibinfo {author} {\bibfnamefont {V.}~\bibnamefont {Goncharov}},
  \bibinfo {author} {\bibfnamefont {T.}~\bibnamefont {Sangster}}, \bibinfo
  {author} {\bibfnamefont {E.}~\bibnamefont {Campbell}}, \bibinfo {author}
  {\bibfnamefont {R.}~\bibnamefont {Betti}}, \bibinfo {author} {\bibfnamefont
  {K.}~\bibnamefont {Anderson}}, \bibinfo {author} {\bibfnamefont
  {T.}~\bibnamefont {Bernat}}, \bibinfo {author} {\bibfnamefont
  {A.}~\bibnamefont {Bose}}, \bibinfo {author} {\bibfnamefont {T.}~\bibnamefont
  {Boehly}}, \bibinfo {author} {\bibfnamefont {M.}~\bibnamefont {Bonino}},
  \emph {et~al.},\ }\bibfield  {title} {\enquote {\bibinfo {title} {The
  national direct-drive program: Omega to the national ignition facility},}\
  }\href@noop {} {\bibfield  {journal} {\bibinfo  {journal} {Fusion Science and
  Technology}\ }\textbf {\bibinfo {volume} {73}},\ \bibinfo {pages} {89--97}
  (\bibinfo {year} {2018})}\BibitemShut {NoStop}%
\bibitem [{\citenamefont {Ceurvorst}\ \emph {et~al.}(2022)\citenamefont
  {Ceurvorst}, \citenamefont {Theobald}, \citenamefont {Rosenberg},
  \citenamefont {Radha}, \citenamefont {Stoeckl}, \citenamefont {Betti},
  \citenamefont {Anderson}, \citenamefont {Marozas}, \citenamefont {Goncharov},
  \citenamefont {Campbell} \emph {et~al.}}]{ceurvorst2022development}%
  \BibitemOpen
  \bibfield  {author} {\bibinfo {author} {\bibfnamefont {L.}~\bibnamefont
  {Ceurvorst}}, \bibinfo {author} {\bibfnamefont {W.}~\bibnamefont {Theobald}},
  \bibinfo {author} {\bibfnamefont {M.}~\bibnamefont {Rosenberg}}, \bibinfo
  {author} {\bibfnamefont {P.}~\bibnamefont {Radha}}, \bibinfo {author}
  {\bibfnamefont {C.}~\bibnamefont {Stoeckl}}, \bibinfo {author} {\bibfnamefont
  {R.}~\bibnamefont {Betti}}, \bibinfo {author} {\bibfnamefont
  {K.}~\bibnamefont {Anderson}}, \bibinfo {author} {\bibfnamefont
  {J.}~\bibnamefont {Marozas}}, \bibinfo {author} {\bibfnamefont
  {V.}~\bibnamefont {Goncharov}}, \bibinfo {author} {\bibfnamefont
  {E.}~\bibnamefont {Campbell}},  \emph {et~al.},\ }\bibfield  {title}
  {\enquote {\bibinfo {title} {Development of an x-ray radiography platform to
  study laser-direct-drive energy coupling at the national ignition
  facility},}\ }\href@noop {} {\bibfield  {journal} {\bibinfo  {journal}
  {Review of Scientific Instruments}\ }\textbf {\bibinfo {volume} {93}}
  (\bibinfo {year} {2022})}\BibitemShut {NoStop}%
\bibitem [{\citenamefont {Viala}\ \emph {et~al.}()\citenamefont {Viala},
  \citenamefont {Cola{\"\i}tis}, \citenamefont {Ceurvorst}, \citenamefont
  {Theobald}, \citenamefont {Igumenshchev}, \citenamefont {Bahukutumbi},
  \citenamefont {Rosenberg}, \citenamefont {Anderson}, \citenamefont {Scott},\
  and\ \citenamefont {Batani}}]{viala2024cbetsolid}%
  \BibitemOpen
  \bibfield  {author} {\bibinfo {author} {\bibfnamefont {D.}~\bibnamefont
  {Viala}}, \bibinfo {author} {\bibfnamefont {A.}~\bibnamefont
  {Cola{\"\i}tis}}, \bibinfo {author} {\bibfnamefont {L.}~\bibnamefont
  {Ceurvorst}}, \bibinfo {author} {\bibfnamefont {W.}~\bibnamefont {Theobald}},
  \bibinfo {author} {\bibfnamefont {I.}~\bibnamefont {Igumenshchev}}, \bibinfo
  {author} {\bibfnamefont {R.}~\bibnamefont {Bahukutumbi}}, \bibinfo {author}
  {\bibfnamefont {M.}~\bibnamefont {Rosenberg}}, \bibinfo {author}
  {\bibfnamefont {K.}~\bibnamefont {Anderson}}, \bibinfo {author}
  {\bibfnamefont {R.}~\bibnamefont {Scott}}, \ and\ \bibinfo {author}
  {\bibfnamefont {D.}~\bibnamefont {Batani}},\ }\href@noop {} {\enquote
  {\bibinfo {title} {Role of cross beam energy transfer in spherical strong
  shock polar direct drive experiments at the national ignition facility},}\
  }\bibinfo {note} {In submission}\BibitemShut {NoStop}%
\bibitem [{\citenamefont {Marozas}\ \emph
  {et~al.}(2018{\natexlab{a}})\citenamefont {Marozas}, \citenamefont
  {Hohenberger}, \citenamefont {Rosenberg}, \citenamefont {Turnbull},
  \citenamefont {Collins}, \citenamefont {Radha}, \citenamefont {McKenty},
  \citenamefont {Zuegel}, \citenamefont {Marshall}, \citenamefont {Regan} \emph
  {et~al.}}]{marozas2018first}%
  \BibitemOpen
  \bibfield  {author} {\bibinfo {author} {\bibfnamefont {J.}~\bibnamefont
  {Marozas}}, \bibinfo {author} {\bibfnamefont {M.}~\bibnamefont
  {Hohenberger}}, \bibinfo {author} {\bibfnamefont {M.}~\bibnamefont
  {Rosenberg}}, \bibinfo {author} {\bibfnamefont {D.}~\bibnamefont {Turnbull}},
  \bibinfo {author} {\bibfnamefont {T.}~\bibnamefont {Collins}}, \bibinfo
  {author} {\bibfnamefont {P.}~\bibnamefont {Radha}}, \bibinfo {author}
  {\bibfnamefont {P.}~\bibnamefont {McKenty}}, \bibinfo {author} {\bibfnamefont
  {J.}~\bibnamefont {Zuegel}}, \bibinfo {author} {\bibfnamefont
  {F.}~\bibnamefont {Marshall}}, \bibinfo {author} {\bibfnamefont
  {S.}~\bibnamefont {Regan}},  \emph {et~al.},\ }\bibfield  {title} {\enquote
  {\bibinfo {title} {First observation of cross-beam energy transfer mitigation
  for direct-drive inertial confinement fusion implosions using wavelength
  detuning at the national ignition facility},}\ }\href@noop {} {\bibfield
  {journal} {\bibinfo  {journal} {Physical review letters}\ }\textbf {\bibinfo
  {volume} {120}},\ \bibinfo {pages} {085001} (\bibinfo {year}
  {2018}{\natexlab{a}})}\BibitemShut {NoStop}%
\bibitem [{\citenamefont {Solodov}\ \emph {et~al.}(2022)\citenamefont
  {Solodov}, \citenamefont {Rosenberg}, \citenamefont {Stoeckl}, \citenamefont
  {Christopherson}, \citenamefont {Betti}, \citenamefont {Radha}, \citenamefont
  {Stoeckl}, \citenamefont {Hohenberger}, \citenamefont {Bachmann},
  \citenamefont {Epstein} \emph {et~al.}}]{solodov2022hot}%
  \BibitemOpen
  \bibfield  {author} {\bibinfo {author} {\bibfnamefont {A.}~\bibnamefont
  {Solodov}}, \bibinfo {author} {\bibfnamefont {M.}~\bibnamefont {Rosenberg}},
  \bibinfo {author} {\bibfnamefont {M.}~\bibnamefont {Stoeckl}}, \bibinfo
  {author} {\bibfnamefont {A.}~\bibnamefont {Christopherson}}, \bibinfo
  {author} {\bibfnamefont {R.}~\bibnamefont {Betti}}, \bibinfo {author}
  {\bibfnamefont {P.}~\bibnamefont {Radha}}, \bibinfo {author} {\bibfnamefont
  {C.}~\bibnamefont {Stoeckl}}, \bibinfo {author} {\bibfnamefont
  {M.}~\bibnamefont {Hohenberger}}, \bibinfo {author} {\bibfnamefont
  {B.}~\bibnamefont {Bachmann}}, \bibinfo {author} {\bibfnamefont
  {R.}~\bibnamefont {Epstein}},  \emph {et~al.},\ }\bibfield  {title} {\enquote
  {\bibinfo {title} {Hot-electron preheat and mitigation in polar-direct-drive
  experiments at the national ignition facility},}\ }\href@noop {} {\bibfield
  {journal} {\bibinfo  {journal} {Physical Review E}\ }\textbf {\bibinfo
  {volume} {106}},\ \bibinfo {pages} {055204} (\bibinfo {year}
  {2022})}\BibitemShut {NoStop}%
\bibitem [{\citenamefont {Barlow}\ \emph {et~al.}(2022)\citenamefont {Barlow},
  \citenamefont {Goffrey}, \citenamefont {Bennett}, \citenamefont {Scott},
  \citenamefont {Glize}, \citenamefont {Theobald}, \citenamefont {Anderson},
  \citenamefont {Solodov}, \citenamefont {Rosenberg}, \citenamefont
  {Hohenberger} \emph {et~al.}}]{barlow2022role}%
  \BibitemOpen
  \bibfield  {author} {\bibinfo {author} {\bibfnamefont {D.}~\bibnamefont
  {Barlow}}, \bibinfo {author} {\bibfnamefont {T.}~\bibnamefont {Goffrey}},
  \bibinfo {author} {\bibfnamefont {K.}~\bibnamefont {Bennett}}, \bibinfo
  {author} {\bibfnamefont {R.}~\bibnamefont {Scott}}, \bibinfo {author}
  {\bibfnamefont {K.}~\bibnamefont {Glize}}, \bibinfo {author} {\bibfnamefont
  {W.}~\bibnamefont {Theobald}}, \bibinfo {author} {\bibfnamefont
  {K.}~\bibnamefont {Anderson}}, \bibinfo {author} {\bibfnamefont
  {A.}~\bibnamefont {Solodov}}, \bibinfo {author} {\bibfnamefont
  {M.}~\bibnamefont {Rosenberg}}, \bibinfo {author} {\bibfnamefont
  {M.}~\bibnamefont {Hohenberger}},  \emph {et~al.},\ }\bibfield  {title}
  {\enquote {\bibinfo {title} {Role of hot electrons in shock ignition
  constrained by experiment at the national ignition facility},}\ }\href@noop
  {} {\bibfield  {journal} {\bibinfo  {journal} {Physics of Plasmas}\ }\textbf
  {\bibinfo {volume} {29}} (\bibinfo {year} {2022})}\BibitemShut {NoStop}%
\bibitem [{\citenamefont {Murphy}\ \emph {et~al.}(2015)\citenamefont {Murphy},
  \citenamefont {Krasheninnikova}, \citenamefont {Kyrala}, \citenamefont
  {Bradley}, \citenamefont {Baumgaertel}, \citenamefont {Cobble}, \citenamefont
  {Hakel}, \citenamefont {Hsu}, \citenamefont {Kline}, \citenamefont
  {Montgomery} \emph {et~al.}}]{murphy2015laser}%
  \BibitemOpen
  \bibfield  {author} {\bibinfo {author} {\bibfnamefont {T.~J.}\ \bibnamefont
  {Murphy}}, \bibinfo {author} {\bibfnamefont {N.~S.}\ \bibnamefont
  {Krasheninnikova}}, \bibinfo {author} {\bibfnamefont {G.}~\bibnamefont
  {Kyrala}}, \bibinfo {author} {\bibfnamefont {P.~A.}\ \bibnamefont {Bradley}},
  \bibinfo {author} {\bibfnamefont {J.~A.}\ \bibnamefont {Baumgaertel}},
  \bibinfo {author} {\bibfnamefont {J.}~\bibnamefont {Cobble}}, \bibinfo
  {author} {\bibfnamefont {P.}~\bibnamefont {Hakel}}, \bibinfo {author}
  {\bibfnamefont {S.~C.}\ \bibnamefont {Hsu}}, \bibinfo {author} {\bibfnamefont
  {J.~L.}\ \bibnamefont {Kline}}, \bibinfo {author} {\bibfnamefont
  {D.}~\bibnamefont {Montgomery}},  \emph {et~al.},\ }\bibfield  {title}
  {\enquote {\bibinfo {title} {Laser irradiance scaling in polar direct drive
  implosions on the national ignition facility},}\ }\href@noop {} {\bibfield
  {journal} {\bibinfo  {journal} {Physics of Plasmas}\ }\textbf {\bibinfo
  {volume} {22}} (\bibinfo {year} {2015})}\BibitemShut {NoStop}%
\bibitem [{\citenamefont {Campbell}\ \emph {et~al.}(2021)\citenamefont
  {Campbell}, \citenamefont {Sangster}, \citenamefont {Goncharov},
  \citenamefont {Zuegel}, \citenamefont {Morse}, \citenamefont {Sorce},
  \citenamefont {Collins}, \citenamefont {Wei}, \citenamefont {Betti},
  \citenamefont {Regan} \emph {et~al.}}]{campbell2021direct}%
  \BibitemOpen
  \bibfield  {author} {\bibinfo {author} {\bibfnamefont {E.}~\bibnamefont
  {Campbell}}, \bibinfo {author} {\bibfnamefont {T.}~\bibnamefont {Sangster}},
  \bibinfo {author} {\bibfnamefont {V.}~\bibnamefont {Goncharov}}, \bibinfo
  {author} {\bibfnamefont {J.}~\bibnamefont {Zuegel}}, \bibinfo {author}
  {\bibfnamefont {S.}~\bibnamefont {Morse}}, \bibinfo {author} {\bibfnamefont
  {C.}~\bibnamefont {Sorce}}, \bibinfo {author} {\bibfnamefont
  {G.}~\bibnamefont {Collins}}, \bibinfo {author} {\bibfnamefont
  {M.}~\bibnamefont {Wei}}, \bibinfo {author} {\bibfnamefont {R.}~\bibnamefont
  {Betti}}, \bibinfo {author} {\bibfnamefont {S.}~\bibnamefont {Regan}},  \emph
  {et~al.},\ }\bibfield  {title} {\enquote {\bibinfo {title} {Direct-drive
  laser fusion: status, plans and future},}\ }\href@noop {} {\bibfield
  {journal} {\bibinfo  {journal} {Philosophical Transactions of the Royal
  Society A}\ }\textbf {\bibinfo {volume} {379}},\ \bibinfo {pages} {20200011}
  (\bibinfo {year} {2021})}\BibitemShut {NoStop}%
\bibitem [{\citenamefont {Rosenberg}\ \emph {et~al.}(2023)\citenamefont
  {Rosenberg}, \citenamefont {Solodov}, \citenamefont {Stoeckl}, \citenamefont
  {Hohenberger}, \citenamefont {Bahukutumbi}, \citenamefont {Theobald},
  \citenamefont {Edgell}, \citenamefont {Filkins}, \citenamefont {Betti},
  \citenamefont {Marshall} \emph {et~al.}}]{rosenberg2023hot}%
  \BibitemOpen
  \bibfield  {author} {\bibinfo {author} {\bibfnamefont {M.}~\bibnamefont
  {Rosenberg}}, \bibinfo {author} {\bibfnamefont {A.}~\bibnamefont {Solodov}},
  \bibinfo {author} {\bibfnamefont {C.}~\bibnamefont {Stoeckl}}, \bibinfo
  {author} {\bibfnamefont {M.}~\bibnamefont {Hohenberger}}, \bibinfo {author}
  {\bibfnamefont {R.}~\bibnamefont {Bahukutumbi}}, \bibinfo {author}
  {\bibfnamefont {W.}~\bibnamefont {Theobald}}, \bibinfo {author}
  {\bibfnamefont {D.}~\bibnamefont {Edgell}}, \bibinfo {author} {\bibfnamefont
  {T.}~\bibnamefont {Filkins}}, \bibinfo {author} {\bibfnamefont
  {R.}~\bibnamefont {Betti}}, \bibinfo {author} {\bibfnamefont
  {F.}~\bibnamefont {Marshall}},  \emph {et~al.},\ }\bibfield  {title}
  {\enquote {\bibinfo {title} {Hot electron preheat in hydrodynamically scaled
  direct-drive inertial confinement fusion implosions on the nif and omega},}\
  }\href@noop {} {\bibfield  {journal} {\bibinfo  {journal} {Physics of
  Plasmas}\ }\textbf {\bibinfo {volume} {30}} (\bibinfo {year}
  {2023})}\BibitemShut {NoStop}%
\bibitem [{\citenamefont {Zylstra}\ \emph {et~al.}(2020)\citenamefont
  {Zylstra}, \citenamefont {Yeamans}, \citenamefont {Le~Pape}, \citenamefont
  {MacKinnon}, \citenamefont {Hohenberger}, \citenamefont {Fittinghoff},
  \citenamefont {Herrmann}, \citenamefont {Kim}, \citenamefont {Radha},
  \citenamefont {McKenty} \emph {et~al.}}]{zylstra2020enhanced}%
  \BibitemOpen
  \bibfield  {author} {\bibinfo {author} {\bibfnamefont {A.}~\bibnamefont
  {Zylstra}}, \bibinfo {author} {\bibfnamefont {C.}~\bibnamefont {Yeamans}},
  \bibinfo {author} {\bibfnamefont {S.}~\bibnamefont {Le~Pape}}, \bibinfo
  {author} {\bibfnamefont {A.}~\bibnamefont {MacKinnon}}, \bibinfo {author}
  {\bibfnamefont {M.}~\bibnamefont {Hohenberger}}, \bibinfo {author}
  {\bibfnamefont {D.}~\bibnamefont {Fittinghoff}}, \bibinfo {author}
  {\bibfnamefont {H.}~\bibnamefont {Herrmann}}, \bibinfo {author}
  {\bibfnamefont {Y.}~\bibnamefont {Kim}}, \bibinfo {author} {\bibfnamefont
  {P.}~\bibnamefont {Radha}}, \bibinfo {author} {\bibfnamefont
  {P.}~\bibnamefont {McKenty}},  \emph {et~al.},\ }\bibfield  {title} {\enquote
  {\bibinfo {title} {Enhanced direct-drive implosion performance on nif with
  wavelength separation},}\ }\href@noop {} {\bibfield  {journal} {\bibinfo
  {journal} {Physics of Plasmas}\ }\textbf {\bibinfo {volume} {27}} (\bibinfo
  {year} {2020})}\BibitemShut {NoStop}%
\bibitem [{\citenamefont {Michel}\ \emph {et~al.}(2009)\citenamefont {Michel},
  \citenamefont {Divol}, \citenamefont {Williams}, \citenamefont {Weber},
  \citenamefont {Thomas}, \citenamefont {Callahan}, \citenamefont {Haan},
  \citenamefont {Salmonson}, \citenamefont {Dixit}, \citenamefont {Hinkel}
  \emph {et~al.}}]{michel2009tuning}%
  \BibitemOpen
  \bibfield  {author} {\bibinfo {author} {\bibfnamefont {P.}~\bibnamefont
  {Michel}}, \bibinfo {author} {\bibfnamefont {L.}~\bibnamefont {Divol}},
  \bibinfo {author} {\bibfnamefont {E.}~\bibnamefont {Williams}}, \bibinfo
  {author} {\bibfnamefont {S.}~\bibnamefont {Weber}}, \bibinfo {author}
  {\bibfnamefont {C.}~\bibnamefont {Thomas}}, \bibinfo {author} {\bibfnamefont
  {D.}~\bibnamefont {Callahan}}, \bibinfo {author} {\bibfnamefont
  {S.}~\bibnamefont {Haan}}, \bibinfo {author} {\bibfnamefont {J.}~\bibnamefont
  {Salmonson}}, \bibinfo {author} {\bibfnamefont {S.}~\bibnamefont {Dixit}},
  \bibinfo {author} {\bibfnamefont {D.}~\bibnamefont {Hinkel}},  \emph
  {et~al.},\ }\bibfield  {title} {\enquote {\bibinfo {title} {Tuning the
  implosion symmetry of icf targets via controlled crossed-beam energy
  transfer},}\ }\href@noop {} {\bibfield  {journal} {\bibinfo  {journal}
  {Physical review letters}\ }\textbf {\bibinfo {volume} {102}},\ \bibinfo
  {pages} {025004} (\bibinfo {year} {2009})}\BibitemShut {NoStop}%
\bibitem [{\citenamefont {Igumenshchev}\ \emph {et~al.}(2010)\citenamefont
  {Igumenshchev}, \citenamefont {Edgell}, \citenamefont {Goncharov},
  \citenamefont {Delettrez}, \citenamefont {Maximov}, \citenamefont {Myatt},
  \citenamefont {Seka}, \citenamefont {Shvydky}, \citenamefont {Skupsky},\ and\
  \citenamefont {Stoeckl}}]{igumenshchev2010crossed}%
  \BibitemOpen
  \bibfield  {author} {\bibinfo {author} {\bibfnamefont {I.}~\bibnamefont
  {Igumenshchev}}, \bibinfo {author} {\bibfnamefont {D.}~\bibnamefont
  {Edgell}}, \bibinfo {author} {\bibfnamefont {V.}~\bibnamefont {Goncharov}},
  \bibinfo {author} {\bibfnamefont {J.}~\bibnamefont {Delettrez}}, \bibinfo
  {author} {\bibfnamefont {A.}~\bibnamefont {Maximov}}, \bibinfo {author}
  {\bibfnamefont {J.}~\bibnamefont {Myatt}}, \bibinfo {author} {\bibfnamefont
  {W.}~\bibnamefont {Seka}}, \bibinfo {author} {\bibfnamefont {A.}~\bibnamefont
  {Shvydky}}, \bibinfo {author} {\bibfnamefont {S.}~\bibnamefont {Skupsky}}, \
  and\ \bibinfo {author} {\bibfnamefont {C.}~\bibnamefont {Stoeckl}},\
  }\bibfield  {title} {\enquote {\bibinfo {title} {Crossed-beam energy transfer
  in implosion experiments on omega},}\ }\href@noop {} {\bibfield  {journal}
  {\bibinfo  {journal} {Physics of Plasmas}\ }\textbf {\bibinfo {volume} {17}}
  (\bibinfo {year} {2010})}\BibitemShut {NoStop}%
\bibitem [{\citenamefont {Marion}\ \emph {et~al.}(2016)\citenamefont {Marion},
  \citenamefont {Debayle}, \citenamefont {Masson-Laborde}, \citenamefont
  {Loiseau},\ and\ \citenamefont {Casanova}}]{marion2016modeling}%
  \BibitemOpen
  \bibfield  {author} {\bibinfo {author} {\bibfnamefont {D.}~\bibnamefont
  {Marion}}, \bibinfo {author} {\bibfnamefont {A.}~\bibnamefont {Debayle}},
  \bibinfo {author} {\bibfnamefont {P.-E.}\ \bibnamefont {Masson-Laborde}},
  \bibinfo {author} {\bibfnamefont {P.}~\bibnamefont {Loiseau}}, \ and\
  \bibinfo {author} {\bibfnamefont {M.}~\bibnamefont {Casanova}},\ }\bibfield
  {title} {\enquote {\bibinfo {title} {Modeling crossed-beam energy transfer
  for inertial confinement fusion},}\ }\href@noop {} {\bibfield  {journal}
  {\bibinfo  {journal} {Physics of Plasmas}\ }\textbf {\bibinfo {volume} {23}}
  (\bibinfo {year} {2016})}\BibitemShut {NoStop}%
\bibitem [{\citenamefont {Cola{\"\i}tis}\ \emph {et~al.}(2021)\citenamefont
  {Cola{\"\i}tis}, \citenamefont {Igumenshchev}, \citenamefont {Mathiaud},\
  and\ \citenamefont {Goncharov}}]{colaitis2021inverse}%
  \BibitemOpen
  \bibfield  {author} {\bibinfo {author} {\bibfnamefont {A.}~\bibnamefont
  {Cola{\"\i}tis}}, \bibinfo {author} {\bibfnamefont {I.}~\bibnamefont
  {Igumenshchev}}, \bibinfo {author} {\bibfnamefont {J.}~\bibnamefont
  {Mathiaud}}, \ and\ \bibinfo {author} {\bibfnamefont {V.}~\bibnamefont
  {Goncharov}},\ }\bibfield  {title} {\enquote {\bibinfo {title} {Inverse ray
  tracing on icosahedral tetrahedron grids for non-linear laser plasma
  interaction coupled to 3d radiation hydrodynamics},}\ }\href@noop {}
  {\bibfield  {journal} {\bibinfo  {journal} {Journal of Computational
  Physics}\ }\textbf {\bibinfo {volume} {443}},\ \bibinfo {pages} {110537}
  (\bibinfo {year} {2021})}\BibitemShut {NoStop}%
\bibitem [{\citenamefont {Follett}\ \emph {et~al.}(2022)\citenamefont
  {Follett}, \citenamefont {Cola{\"\i}tis}, \citenamefont {Turnbull},
  \citenamefont {Froula},\ and\ \citenamefont
  {Palastro}}]{follett2022validation}%
  \BibitemOpen
  \bibfield  {author} {\bibinfo {author} {\bibfnamefont {R.}~\bibnamefont
  {Follett}}, \bibinfo {author} {\bibfnamefont {A.}~\bibnamefont
  {Cola{\"\i}tis}}, \bibinfo {author} {\bibfnamefont {D.}~\bibnamefont
  {Turnbull}}, \bibinfo {author} {\bibfnamefont {D.}~\bibnamefont {Froula}}, \
  and\ \bibinfo {author} {\bibfnamefont {J.}~\bibnamefont {Palastro}},\
  }\bibfield  {title} {\enquote {\bibinfo {title} {Validation of ray-based
  cross-beam energy transfer models},}\ }\href@noop {} {\bibfield  {journal}
  {\bibinfo  {journal} {Physics of Plasmas}\ }\textbf {\bibinfo {volume} {29}}
  (\bibinfo {year} {2022})}\BibitemShut {NoStop}%
\bibitem [{\citenamefont {Turnbull}\ \emph {et~al.}(2020)\citenamefont
  {Turnbull}, \citenamefont {Cola{\"\i}tis}, \citenamefont {Hansen},
  \citenamefont {Milder}, \citenamefont {Palastro}, \citenamefont {Katz},
  \citenamefont {Dorrer}, \citenamefont {Kruschwitz}, \citenamefont {Strozzi},\
  and\ \citenamefont {Froula}}]{turnbull2020impact}%
  \BibitemOpen
  \bibfield  {author} {\bibinfo {author} {\bibfnamefont {D.}~\bibnamefont
  {Turnbull}}, \bibinfo {author} {\bibfnamefont {A.}~\bibnamefont
  {Cola{\"\i}tis}}, \bibinfo {author} {\bibfnamefont {A.~M.}\ \bibnamefont
  {Hansen}}, \bibinfo {author} {\bibfnamefont {A.~L.}\ \bibnamefont {Milder}},
  \bibinfo {author} {\bibfnamefont {J.~P.}\ \bibnamefont {Palastro}}, \bibinfo
  {author} {\bibfnamefont {J.}~\bibnamefont {Katz}}, \bibinfo {author}
  {\bibfnamefont {C.}~\bibnamefont {Dorrer}}, \bibinfo {author} {\bibfnamefont
  {B.~E.}\ \bibnamefont {Kruschwitz}}, \bibinfo {author} {\bibfnamefont
  {D.~J.}\ \bibnamefont {Strozzi}}, \ and\ \bibinfo {author} {\bibfnamefont
  {D.~H.}\ \bibnamefont {Froula}},\ }\bibfield  {title} {\enquote {\bibinfo
  {title} {Impact of the langdon effect on crossed-beam energy transfer},}\
  }\href@noop {} {\bibfield  {journal} {\bibinfo  {journal} {Nature Physics}\
  }\textbf {\bibinfo {volume} {16}},\ \bibinfo {pages} {181--185} (\bibinfo
  {year} {2020})}\BibitemShut {NoStop}%
\bibitem [{\citenamefont {Hansen}\ \emph {et~al.}(2021)\citenamefont {Hansen},
  \citenamefont {Nguyen}, \citenamefont {Turnbull}, \citenamefont {Albright},
  \citenamefont {Follett}, \citenamefont {Huff}, \citenamefont {Katz},
  \citenamefont {Mastrosimone}, \citenamefont {Milder}, \citenamefont {Yin}
  \emph {et~al.}}]{hansen2021cross}%
  \BibitemOpen
  \bibfield  {author} {\bibinfo {author} {\bibfnamefont {A.}~\bibnamefont
  {Hansen}}, \bibinfo {author} {\bibfnamefont {K.}~\bibnamefont {Nguyen}},
  \bibinfo {author} {\bibfnamefont {D.}~\bibnamefont {Turnbull}}, \bibinfo
  {author} {\bibfnamefont {B.}~\bibnamefont {Albright}}, \bibinfo {author}
  {\bibfnamefont {R.}~\bibnamefont {Follett}}, \bibinfo {author} {\bibfnamefont
  {R.}~\bibnamefont {Huff}}, \bibinfo {author} {\bibfnamefont {J.}~\bibnamefont
  {Katz}}, \bibinfo {author} {\bibfnamefont {D.}~\bibnamefont {Mastrosimone}},
  \bibinfo {author} {\bibfnamefont {A.}~\bibnamefont {Milder}}, \bibinfo
  {author} {\bibfnamefont {L.}~\bibnamefont {Yin}},  \emph {et~al.},\
  }\bibfield  {title} {\enquote {\bibinfo {title} {Cross-beam energy transfer
  saturation by ion heating},}\ }\href@noop {} {\bibfield  {journal} {\bibinfo
  {journal} {Physical Review Letters}\ }\textbf {\bibinfo {volume} {126}},\
  \bibinfo {pages} {075002} (\bibinfo {year} {2021})}\BibitemShut {NoStop}%
\bibitem [{\citenamefont {Manheimer}, \citenamefont {Colombant},\ and\
  \citenamefont {Gardner}(1982)}]{manheimer1982steady}%
  \BibitemOpen
  \bibfield  {author} {\bibinfo {author} {\bibfnamefont {W.~M.}\ \bibnamefont
  {Manheimer}}, \bibinfo {author} {\bibfnamefont {D.}~\bibnamefont
  {Colombant}}, \ and\ \bibinfo {author} {\bibfnamefont {J.}~\bibnamefont
  {Gardner}},\ }\bibfield  {title} {\enquote {\bibinfo {title} {Steady-state
  planar ablative flow},}\ }\href@noop {} {\bibfield  {journal} {\bibinfo
  {journal} {The Physics of Fluids}\ }\textbf {\bibinfo {volume} {25}},\
  \bibinfo {pages} {1644--1652} (\bibinfo {year} {1982})}\BibitemShut {NoStop}%
\bibitem [{\citenamefont {Scheiner}\ and\ \citenamefont
  {Schmitt}(2019)}]{scheiner2019role}%
  \BibitemOpen
  \bibfield  {author} {\bibinfo {author} {\bibfnamefont {B.}~\bibnamefont
  {Scheiner}}\ and\ \bibinfo {author} {\bibfnamefont {M.}~\bibnamefont
  {Schmitt}},\ }\bibfield  {title} {\enquote {\bibinfo {title} {The role of
  incidence angle in the laser ablation of a planar target},}\ }\href@noop {}
  {\bibfield  {journal} {\bibinfo  {journal} {Physics of Plasmas}\ }\textbf
  {\bibinfo {volume} {26}} (\bibinfo {year} {2019})}\BibitemShut {NoStop}%
\bibitem [{\citenamefont {Murakami}(1995)}]{murakami1995irradiation}%
  \BibitemOpen
  \bibfield  {author} {\bibinfo {author} {\bibfnamefont {M.}~\bibnamefont
  {Murakami}},\ }\bibfield  {title} {\enquote {\bibinfo {title} {Irradiation
  system based on dodecahedron for inertial confinement fusion},}\ }\href@noop
  {} {\bibfield  {journal} {\bibinfo  {journal} {Applied physics letters}\
  }\textbf {\bibinfo {volume} {66}},\ \bibinfo {pages} {1587--1589} (\bibinfo
  {year} {1995})}\BibitemShut {NoStop}%
\bibitem [{\citenamefont {Murakami}\ \emph {et~al.}(2010)\citenamefont
  {Murakami}, \citenamefont {Sarukura}, \citenamefont {Azechi}, \citenamefont
  {Temporal},\ and\ \citenamefont {Schmitt}}]{murakami2010optimization}%
  \BibitemOpen
  \bibfield  {author} {\bibinfo {author} {\bibfnamefont {M.}~\bibnamefont
  {Murakami}}, \bibinfo {author} {\bibfnamefont {N.}~\bibnamefont {Sarukura}},
  \bibinfo {author} {\bibfnamefont {H.}~\bibnamefont {Azechi}}, \bibinfo
  {author} {\bibfnamefont {M.}~\bibnamefont {Temporal}}, \ and\ \bibinfo
  {author} {\bibfnamefont {A.}~\bibnamefont {Schmitt}},\ }\bibfield  {title}
  {\enquote {\bibinfo {title} {Optimization of irradiation configuration in
  laser fusion utilizing self-organizing electrodynamic system},}\ }\href@noop
  {} {\bibfield  {journal} {\bibinfo  {journal} {Physics of Plasmas}\ }\textbf
  {\bibinfo {volume} {17}} (\bibinfo {year} {2010})}\BibitemShut {NoStop}%
\bibitem [{\citenamefont {Shvydky}\ \emph {et~al.}(2022)\citenamefont
  {Shvydky}, \citenamefont {Trickey}, \citenamefont {Maximov}, \citenamefont
  {Igumenshchev}, \citenamefont {McKenty},\ and\ \citenamefont
  {Goncharov}}]{shvydky2022optimization}%
  \BibitemOpen
  \bibfield  {author} {\bibinfo {author} {\bibfnamefont {A.}~\bibnamefont
  {Shvydky}}, \bibinfo {author} {\bibfnamefont {W.}~\bibnamefont {Trickey}},
  \bibinfo {author} {\bibfnamefont {A.}~\bibnamefont {Maximov}}, \bibinfo
  {author} {\bibfnamefont {I.}~\bibnamefont {Igumenshchev}}, \bibinfo {author}
  {\bibfnamefont {P.}~\bibnamefont {McKenty}}, \ and\ \bibinfo {author}
  {\bibfnamefont {V.}~\bibnamefont {Goncharov}},\ }\bibfield  {title} {\enquote
  {\bibinfo {title} {Optimization of irradiation configuration using spherical
  t-designs for laser-direct-drive inertial confinement fusion},}\ }\href@noop
  {} {\bibfield  {journal} {\bibinfo  {journal} {Nuclear Fusion}\ }\textbf
  {\bibinfo {volume} {63}},\ \bibinfo {pages} {014004} (\bibinfo {year}
  {2022})}\BibitemShut {NoStop}%
\bibitem [{\citenamefont {Hohenberger}\ \emph {et~al.}(2015)\citenamefont
  {Hohenberger}, \citenamefont {Radha}, \citenamefont {Myatt}, \citenamefont
  {LePape}, \citenamefont {Marozas}, \citenamefont {Marshall}, \citenamefont
  {Michel}, \citenamefont {Regan}, \citenamefont {Seka}, \citenamefont
  {Shvydky} \emph {et~al.}}]{hohenberger2015polar}%
  \BibitemOpen
  \bibfield  {author} {\bibinfo {author} {\bibfnamefont {M.}~\bibnamefont
  {Hohenberger}}, \bibinfo {author} {\bibfnamefont {P.}~\bibnamefont {Radha}},
  \bibinfo {author} {\bibfnamefont {J.}~\bibnamefont {Myatt}}, \bibinfo
  {author} {\bibfnamefont {S.}~\bibnamefont {LePape}}, \bibinfo {author}
  {\bibfnamefont {J.}~\bibnamefont {Marozas}}, \bibinfo {author} {\bibfnamefont
  {F.}~\bibnamefont {Marshall}}, \bibinfo {author} {\bibfnamefont
  {D.}~\bibnamefont {Michel}}, \bibinfo {author} {\bibfnamefont
  {S.}~\bibnamefont {Regan}}, \bibinfo {author} {\bibfnamefont
  {W.}~\bibnamefont {Seka}}, \bibinfo {author} {\bibfnamefont {A.}~\bibnamefont
  {Shvydky}},  \emph {et~al.},\ }\bibfield  {title} {\enquote {\bibinfo {title}
  {Polar-direct-drive experiments on the national ignition facility},}\
  }\href@noop {} {\bibfield  {journal} {\bibinfo  {journal} {Physics of
  Plasmas}\ }\textbf {\bibinfo {volume} {22}} (\bibinfo {year}
  {2015})}\BibitemShut {NoStop}%
\bibitem [{\citenamefont {Collins}\ and\ \citenamefont
  {Marozas}(2018)}]{collins2018mitigation}%
  \BibitemOpen
  \bibfield  {author} {\bibinfo {author} {\bibfnamefont {T.}~\bibnamefont
  {Collins}}\ and\ \bibinfo {author} {\bibfnamefont {J.}~\bibnamefont
  {Marozas}},\ }\bibfield  {title} {\enquote {\bibinfo {title} {Mitigation of
  cross-beam energy transfer in ignition-scale polar-direct-drive target
  designs for the national ignition facility},}\ }\href@noop {} {\bibfield
  {journal} {\bibinfo  {journal} {Physics of Plasmas}\ }\textbf {\bibinfo
  {volume} {25}} (\bibinfo {year} {2018})}\BibitemShut {NoStop}%
\bibitem [{\citenamefont {Marozas}\ \emph
  {et~al.}(2018{\natexlab{b}})\citenamefont {Marozas}, \citenamefont
  {Hohenberger}, \citenamefont {Rosenberg}, \citenamefont {Turnbull},
  \citenamefont {Collins}, \citenamefont {Radha}, \citenamefont {McKenty},
  \citenamefont {Zuegel}, \citenamefont {Marshall}, \citenamefont {Regan} \emph
  {et~al.}}]{marozas2018wavelength}%
  \BibitemOpen
  \bibfield  {author} {\bibinfo {author} {\bibfnamefont {J.}~\bibnamefont
  {Marozas}}, \bibinfo {author} {\bibfnamefont {M.}~\bibnamefont
  {Hohenberger}}, \bibinfo {author} {\bibfnamefont {M.}~\bibnamefont
  {Rosenberg}}, \bibinfo {author} {\bibfnamefont {D.}~\bibnamefont {Turnbull}},
  \bibinfo {author} {\bibfnamefont {T.}~\bibnamefont {Collins}}, \bibinfo
  {author} {\bibfnamefont {P.}~\bibnamefont {Radha}}, \bibinfo {author}
  {\bibfnamefont {P.}~\bibnamefont {McKenty}}, \bibinfo {author} {\bibfnamefont
  {J.}~\bibnamefont {Zuegel}}, \bibinfo {author} {\bibfnamefont
  {F.}~\bibnamefont {Marshall}}, \bibinfo {author} {\bibfnamefont
  {S.}~\bibnamefont {Regan}},  \emph {et~al.},\ }\bibfield  {title} {\enquote
  {\bibinfo {title} {Wavelength-detuning cross-beam energy transfer mitigation
  scheme for direct drive: Modeling and evidence from national ignition
  facility implosions},}\ }\href@noop {} {\bibfield  {journal} {\bibinfo
  {journal} {Physics of Plasmas}\ }\textbf {\bibinfo {volume} {25}} (\bibinfo
  {year} {2018}{\natexlab{b}})}\BibitemShut {NoStop}%
\bibitem [{\citenamefont {Cola{\"\i}tis}\ \emph {et~al.}(2019)\citenamefont
  {Cola{\"\i}tis}, \citenamefont {Follett}, \citenamefont {Palastro},
  \citenamefont {Igumenschev},\ and\ \citenamefont
  {Goncharov}}]{colaitis2019adaptive}%
  \BibitemOpen
  \bibfield  {author} {\bibinfo {author} {\bibfnamefont {A.}~\bibnamefont
  {Cola{\"\i}tis}}, \bibinfo {author} {\bibfnamefont {R.}~\bibnamefont
  {Follett}}, \bibinfo {author} {\bibfnamefont {J.}~\bibnamefont {Palastro}},
  \bibinfo {author} {\bibfnamefont {I.}~\bibnamefont {Igumenschev}}, \ and\
  \bibinfo {author} {\bibfnamefont {V.}~\bibnamefont {Goncharov}},\ }\bibfield
  {title} {\enquote {\bibinfo {title} {Adaptive inverse ray-tracing for
  accurate and efficient modeling of cross beam energy transfer in
  hydrodynamics simulations},}\ }\href@noop {} {\bibfield  {journal} {\bibinfo
  {journal} {Physics of Plasmas}\ }\textbf {\bibinfo {volume} {26}} (\bibinfo
  {year} {2019})}\BibitemShut {NoStop}%
\bibitem [{\citenamefont {Igumenshchev}\ \emph {et~al.}(2016)\citenamefont
  {Igumenshchev}, \citenamefont {Goncharov}, \citenamefont {Marshall},
  \citenamefont {Knauer}, \citenamefont {Campbell}, \citenamefont {Forrest},
  \citenamefont {Froula}, \citenamefont {Glebov}, \citenamefont {McCrory},
  \citenamefont {Regan} \emph {et~al.}}]{igumenshchev2016three}%
  \BibitemOpen
  \bibfield  {author} {\bibinfo {author} {\bibfnamefont {I.}~\bibnamefont
  {Igumenshchev}}, \bibinfo {author} {\bibfnamefont {V.}~\bibnamefont
  {Goncharov}}, \bibinfo {author} {\bibfnamefont {F.}~\bibnamefont {Marshall}},
  \bibinfo {author} {\bibfnamefont {J.}~\bibnamefont {Knauer}}, \bibinfo
  {author} {\bibfnamefont {E.}~\bibnamefont {Campbell}}, \bibinfo {author}
  {\bibfnamefont {C.}~\bibnamefont {Forrest}}, \bibinfo {author} {\bibfnamefont
  {D.}~\bibnamefont {Froula}}, \bibinfo {author} {\bibfnamefont {V.~Y.}\
  \bibnamefont {Glebov}}, \bibinfo {author} {\bibfnamefont {R.}~\bibnamefont
  {McCrory}}, \bibinfo {author} {\bibfnamefont {S.}~\bibnamefont {Regan}},
  \emph {et~al.},\ }\bibfield  {title} {\enquote {\bibinfo {title}
  {Three-dimensional modeling of direct-drive cryogenic implosions on omega},}\
  }\href@noop {} {\bibfield  {journal} {\bibinfo  {journal} {Physics of
  Plasmas}\ }\textbf {\bibinfo {volume} {23}} (\bibinfo {year}
  {2016})}\BibitemShut {NoStop}%
\bibitem [{\citenamefont {Keane}(2014)}]{keane2014national}%
  \BibitemOpen
  \bibfield  {author} {\bibinfo {author} {\bibfnamefont {C.}~\bibnamefont
  {Keane}},\ }\href@noop {} {\enquote {\bibinfo {title} {National ignition
  facility user guide},}\ }\bibinfo {type} {Tech. Rep.}\ (\bibinfo
  {institution} {Lawrence Livermore National Lab.(LLNL), Livermore, CA (United
  States)},\ \bibinfo {year} {2014})\BibitemShut {NoStop}%
\bibitem [{\citenamefont {Kyrala}\ \emph {et~al.}(2010)\citenamefont {Kyrala},
  \citenamefont {Dixit}, \citenamefont {Glenzer}, \citenamefont {Kalantar},
  \citenamefont {Bradley}, \citenamefont {Izumi}, \citenamefont {Meezan},
  \citenamefont {Landen}, \citenamefont {Callahan}, \citenamefont {Weber} \emph
  {et~al.}}]{kyrala2010gxd}%
  \BibitemOpen
  \bibfield  {author} {\bibinfo {author} {\bibfnamefont {G.}~\bibnamefont
  {Kyrala}}, \bibinfo {author} {\bibfnamefont {S.}~\bibnamefont {Dixit}},
  \bibinfo {author} {\bibfnamefont {S.}~\bibnamefont {Glenzer}}, \bibinfo
  {author} {\bibfnamefont {D.}~\bibnamefont {Kalantar}}, \bibinfo {author}
  {\bibfnamefont {D.}~\bibnamefont {Bradley}}, \bibinfo {author} {\bibfnamefont
  {N.}~\bibnamefont {Izumi}}, \bibinfo {author} {\bibfnamefont
  {N.}~\bibnamefont {Meezan}}, \bibinfo {author} {\bibfnamefont
  {O.}~\bibnamefont {Landen}}, \bibinfo {author} {\bibfnamefont
  {D.}~\bibnamefont {Callahan}}, \bibinfo {author} {\bibfnamefont
  {S.}~\bibnamefont {Weber}},  \emph {et~al.},\ }\bibfield  {title} {\enquote
  {\bibinfo {title} {Measuring symmetry of implosions in cryogenic hohlraums at
  the nif using gated x-ray detectors},}\ }\href@noop {} {\bibfield  {journal}
  {\bibinfo  {journal} {Review of Scientific Instruments}\ }\textbf {\bibinfo
  {volume} {81}},\ \bibinfo {pages} {10E316} (\bibinfo {year}
  {2010})}\BibitemShut {NoStop}%
\bibitem [{\citenamefont {Gad}(2021)}]{gad2021pygad}%
  \BibitemOpen
  \bibfield  {author} {\bibinfo {author} {\bibfnamefont {A.~F.}\ \bibnamefont
  {Gad}},\ }\bibfield  {title} {\enquote {\bibinfo {title} {Pygad: An intuitive
  genetic algorithm python library},}\ }\href@noop {} {\bibfield  {journal}
  {\bibinfo  {journal} {arXiv preprint arXiv:2106.06158}\ } (\bibinfo {year}
  {2021})}\BibitemShut {NoStop}%
\bibitem [{\citenamefont {Wright}(2015)}]{wright2015coordinate}%
  \BibitemOpen
  \bibfield  {author} {\bibinfo {author} {\bibfnamefont {S.~J.}\ \bibnamefont
  {Wright}},\ }\bibfield  {title} {\enquote {\bibinfo {title} {Coordinate
  descent algorithms},}\ }\href@noop {} {\bibfield  {journal} {\bibinfo
  {journal} {Mathematical programming}\ }\textbf {\bibinfo {volume} {151}},\
  \bibinfo {pages} {3--34} (\bibinfo {year} {2015})}\BibitemShut {NoStop}%
\bibitem [{\citenamefont {Epstein}(2004)}]{epstein2004bell}%
  \BibitemOpen
  \bibfield  {author} {\bibinfo {author} {\bibfnamefont {R.}~\bibnamefont
  {Epstein}},\ }\bibfield  {title} {\enquote {\bibinfo {title} {On the
  bell--plesset effects: the effects of uniform compression and geometrical
  convergence on the classical rayleigh--taylor instability},}\ }\href@noop {}
  {\bibfield  {journal} {\bibinfo  {journal} {Physics of plasmas}\ }\textbf
  {\bibinfo {volume} {11}},\ \bibinfo {pages} {5114--5124} (\bibinfo {year}
  {2004})}\BibitemShut {NoStop}%
\bibitem [{\citenamefont {Bro}\ and\ \citenamefont
  {Smilde}(2014)}]{bro2014principal}%
  \BibitemOpen
  \bibfield  {author} {\bibinfo {author} {\bibfnamefont {R.}~\bibnamefont
  {Bro}}\ and\ \bibinfo {author} {\bibfnamefont {A.~K.}\ \bibnamefont
  {Smilde}},\ }\bibfield  {title} {\enquote {\bibinfo {title} {Principal
  component analysis},}\ }\href@noop {} {\bibfield  {journal} {\bibinfo
  {journal} {Analytical methods}\ }\textbf {\bibinfo {volume} {6}},\ \bibinfo
  {pages} {2812--2831} (\bibinfo {year} {2014})}\BibitemShut {NoStop}%
\bibitem [{\citenamefont {Cola{\"\i}tis}\ \emph {et~al.}(2023)\citenamefont
  {Cola{\"\i}tis}, \citenamefont {Follett}, \citenamefont {Dorrer},
  \citenamefont {Seaton}, \citenamefont {Viala}, \citenamefont {Igumenshchev},
  \citenamefont {Turnbull}, \citenamefont {Goncharov},\ and\ \citenamefont
  {Froula}}]{colaitis2023exploration}%
  \BibitemOpen
  \bibfield  {author} {\bibinfo {author} {\bibfnamefont {A.}~\bibnamefont
  {Cola{\"\i}tis}}, \bibinfo {author} {\bibfnamefont {R.~K.}\ \bibnamefont
  {Follett}}, \bibinfo {author} {\bibfnamefont {C.}~\bibnamefont {Dorrer}},
  \bibinfo {author} {\bibfnamefont {A.~G.}\ \bibnamefont {Seaton}}, \bibinfo
  {author} {\bibfnamefont {D.}~\bibnamefont {Viala}}, \bibinfo {author}
  {\bibfnamefont {I.}~\bibnamefont {Igumenshchev}}, \bibinfo {author}
  {\bibfnamefont {D.}~\bibnamefont {Turnbull}}, \bibinfo {author}
  {\bibfnamefont {V.}~\bibnamefont {Goncharov}}, \ and\ \bibinfo {author}
  {\bibfnamefont {D.~H.}\ \bibnamefont {Froula}},\ }\bibfield  {title}
  {\enquote {\bibinfo {title} {Exploration of cross-beam energy transfer
  mitigation constraints for designing an ignition-scale direct-drive inertial
  confinement fusion driver},}\ }\href@noop {} {\bibfield  {journal} {\bibinfo
  {journal} {Physics of Plasmas}\ }\textbf {\bibinfo {volume} {30}} (\bibinfo
  {year} {2023})}\BibitemShut {NoStop}%
\bibitem [{\citenamefont {Eimerl}\ \emph {et~al.}(2014)\citenamefont {Eimerl},
  \citenamefont {Campbell}, \citenamefont {Krupke}, \citenamefont {Zweiback},
  \citenamefont {Kruer}, \citenamefont {Marozas}, \citenamefont {Zuegel},
  \citenamefont {Myatt}, \citenamefont {Kelly}, \citenamefont {Froula} \emph
  {et~al.}}]{eimerl2014stardriver}%
  \BibitemOpen
  \bibfield  {author} {\bibinfo {author} {\bibfnamefont {D.}~\bibnamefont
  {Eimerl}}, \bibinfo {author} {\bibfnamefont {E.~M.}\ \bibnamefont
  {Campbell}}, \bibinfo {author} {\bibfnamefont {W.~F.}\ \bibnamefont
  {Krupke}}, \bibinfo {author} {\bibfnamefont {J.}~\bibnamefont {Zweiback}},
  \bibinfo {author} {\bibfnamefont {W.}~\bibnamefont {Kruer}}, \bibinfo
  {author} {\bibfnamefont {J.}~\bibnamefont {Marozas}}, \bibinfo {author}
  {\bibfnamefont {J.}~\bibnamefont {Zuegel}}, \bibinfo {author} {\bibfnamefont
  {J.}~\bibnamefont {Myatt}}, \bibinfo {author} {\bibfnamefont
  {J.}~\bibnamefont {Kelly}}, \bibinfo {author} {\bibfnamefont
  {D.}~\bibnamefont {Froula}},  \emph {et~al.},\ }\bibfield  {title} {\enquote
  {\bibinfo {title} {Stardriver: a flexible laser driver for inertial
  confinement fusion and high energy density physics},}\ }\href@noop {}
  {\bibfield  {journal} {\bibinfo  {journal} {Journal of Fusion Energy}\
  }\textbf {\bibinfo {volume} {33}},\ \bibinfo {pages} {476--488} (\bibinfo
  {year} {2014})}\BibitemShut {NoStop}%
\end{thebibliography}%

\end{document}